\documentclass[letterpaper]{article}
\usepackage[english]{babel}
\usepackage[utf8x]{inputenc}
\usepackage[T1]{fontenc}
\usepackage{amsmath}
\usepackage{amssymb}
\usepackage{amsthm}
\usepackage{graphicx}
\usepackage[margin=1in]{geometry}
\usepackage{float}
 \usepackage[linesnumbered,lined,boxed]{algorithm2e}
\usepackage{algpseudocode}
\usepackage{hyperref}
\usepackage{listings}
\usepackage{pdflscape}
\usepackage{setspace}
\usepackage{longtable}
\usepackage{young}
\usepackage{tikz-cd}
\usepackage{thmtools}
\usepackage[affil-it]{authblk}
\usepackage[english]{babel}
\usepackage{blindtext}
\usepackage{multicol}
\usepackage{authblk}
\usepackage{qcircuit}
\usepackage{rotating}
\usepackage{bbm}
\usepackage{cite}
\usepackage{caption}
\usepackage{subcaption}
\usepackage{mhchem}

\definecolor{IKblue}{RGB}{0,0,100}

\title{Noise reduction using past causal cones \\ 
in variational quantum algorithms}

\author[1]{Omar Shehab\footnote{Electronic address: shehab@ionq.co}}
\author[2]{Isaac H. Kim\footnote{Electronic address: isaac.kim.quantum@gmail.com}}
\author[3]{Nhung H. Nguyen}
\author[3]{Kevin Landsman}
\author[3,4]{\\Cinthia H. Alderete}
\author[3]{Daiwei Zhu}
\author[1,3]{C. Monroe}
\author[3]{Norbert M. Linke}
\affil[1]{\small IonQ, Inc., College Park, MD 20740}
\affil[2]{Stanford Institute for Theoretical Physics, Stanford University, Stanford CA 94305}
\affil[3]{Joint Quantum Institute, Department of Physics and Joint Center for Quantum Information and Computer Science, University of Maryland, College Park, MD 20742}
\affil[4]{Instituto Nacional de Astrof\'{i}sica, \'{O}ptica y Electr\'{o}nica, Calle Luis 
Enrique Erro No. 1, Sta. Ma. Tonantzintla, Pue. CP 72840, Mexico}

\begin{document}

\maketitle

\begin{abstract}
We introduce an approach to improve the accuracy and reduce the sample complexity of near term quantum-classical algorithms. We construct a simpler initial parameterized quantum state, or ansatz, based on the past causal cone of each observable, generally yielding fewer qubits and gates. We implement this protocol on a trapped ion quantum computer and demonstrate improvement in accuracy and time-to-solution at an arbitrary point in the variational search space. We report a $\sim 27\%$ improvement in the accuracy of the  calculation of the deuteron binding energy and $\sim 40\%$ improvement in the accuracy of the quantum approximate optimization of the MAXCUT problem applied to the dragon graph $T_{3,2}$. When the time-to-solution is prioritized over accuracy, the former requires $\sim 71\%$ fewer measurements and the latter requires $\sim 78\%$ fewer measurements.

\end{abstract}


\section{Introduction}
The variational quantum eigensolver algorithm (VQE) \cite{peruzzo2014variational, mcclean2016theory} has been proposed and demonstrated for eigenvalue approximation problems on noisy intermediate-scale quantum (NISQ) computers \cite{preskill2018quantum}. The VQE algorithm off-loads part of the task onto a classical computer in a hybrid quantum-classical approach with short-depth quantum circuits, as opposed to the more stringent gate fidelity requirements in the phase estimation algorithm approach \cite{kitaev1995quantum, dobvsivcek2007arbitrary}. Researchers have successfully implemented the algorithm on various quantum hardware \cite{peruzzo2014variational, o2016scalable, shen2017quantum, kandala2017hardware, colless2018computation, santagati2018witnessing, hempel2018quantum, dumitrescu2018cloud, klco2018quantum, nam2019ground}, also showing that the VQE algorithm is robust to certain types of error \cite{kandala2017hardware, mcclean2016theory}.

The quantum approximate optimization algorithm (QAOA) \cite{farhi2014quantum} has been proposed to solve combinatorial optimization problems on a NISQ computer. While the domain of application and the details of QAOA implementation differ significantly from VQE algorithms, from a high level point of view, these approaches are  similar in nature. In this work, we focus on noise reduction techniques in VQE algorithms, but the discussions, experiments, and results are equally pertinent to both VQE and QAOA algorithms.

VQE uses the Rayleigh-Ritz variational principle \cite{macdonald1934modified, weinstein1934modified} to compute the eigenvalue of a Hamiltonian $H$. For a parameterized wavefunction $\Psi(\vec{\theta})$, the energy expectation $\langle \Psi(\vec{\theta})| H  |\Psi(\vec{\theta})\rangle$ is bounded from below by the lowest eigenvalue $E_0$ of the Hamiltonian, where $\vec{\theta}$ is a vector of independent parameters. VQE relies on the efficient creation of candidate states $|\Psi(\vec{\theta})\rangle$ and the measurement of $\langle \Psi(\vec{\theta})| H  |\Psi(\vec{\theta})\rangle$ using a quantum computer. By classically optimizing the parameters $\vec{\theta}$, the local minimum of the Hamiltonian cost function is taken as an approximate ground state energy $E_0$ of the system. QAOAs arrive at a target state by applying $p$ layers of evolution. While increasing the total number of gates and variational parameters, each successive layer refines the candidate state and improves the accuracy of the approximation.

To generate the parameterized wavefunction for VQE/QAOA, both a hardware efficient ansatz \cite{kandala2017hardware} and a physically inspired ansatz \cite{dumitrescu2018cloud, barkoutsos2018quantum} have been implemented with respective advantages and disadvantages.  The hardware efficient ansatz \cite{barkoutsos2018quantum} suffers from the potential for getting stuck in the barren plateaus of the energy landscape \cite{mcclean2018barren}. The physically inspired ansatz can quickly lead to deep circuits as the complexity of the physical system increases. For example, for the unitary coupled cluster (UCC) ansatz relevant to molecular simulations, the numbers of gates and circuit depth scale as $O\left(M^3 N\right)$ and $O\left(M^2 N\right)$ respectively where $M$ is the number of spin-orbitals, and $N$ is the number of electrons, assuming a single Trotter step \cite{motta2018low, o2019generalized}. Finally, the success of VQE also depends on a large number of measurements for statistical certainty.

The coefficient of the Pauli term with the largest absolute value in a qubit Hamiltonian determines the upper bound on the variance of the expectation value \cite{kandala2017hardware} and hence the hardware performance and number of measurements needed to achieve a desired accuracy. It can be limited by a careful model choice, as done e.g. in \cite{shehab2019toward} when computing the binding energy of the deuteron nucleus. Once an appropriate fermionic model is constructed for the VQE algorithm,  the accuracy of the result is determined by the number of measurements and experimental details such as the gate fidelity and qubit connectivity. Clearly, any reduction in circuit complexity, the size of the parameter space, or the number of measurements is desirable for a successful VQE application.

Several approaches to optimize VQE circuits have been proposed, such as removing qubits stabilized by the Hamiltonian \cite{o2016scalable}, making use of block-diagonality \cite{moll2016optimizing} or symmetry \cite{bravyi2017tapering, nam2019ground}, grouping Hamiltonian terms based on their norms \cite{hadfield2018divide}, Pauli grouping \cite{kandala2017hardware}, resetting qubits in a tensor network representation \cite{peng2019simulating, 1902.02663}, or subspace expansion \cite{takeshita2019increasing}. The effect of optimization on VQE accuracy has also been rigorously studied \cite{ kandala2017hardware}. However, choosing the appropriate values for the QAOA circuit parameters to reach global optima has been shown to be a hard problem \cite{mcclean2018barren}. In this work, we improve the VQE/QAOA process fidelity by using reduced ansatz circuits based on past causal cones of each observable, and experimentally demonstrate the advantage on a trapped ion quantum computer.

\section{Past causal cones as a reduced variational ansatz}
\label{sec:nra}
The reduced-ansatz variational quantum eigensolver (RA-VQE) algorithm (and similarly the reduced-ansatz quantum approximate optimization algorithm, RA-QAOA) leverages the construction of a reduced ansatz with respect to the terms of the Hamiltonian. Our construction shares similarities to the deep multi-scale entanglement renormalization
ansatz (DMERA) proposed in \cite{kim2017robust, kim2017noise}, but can be applied more generally. The algorithm replaces the original ansatz with a set of reduced circuits computed from the past causal cone (PCC) \cite{evenbly2009algorithms} of each term in the Hamiltonian. The PCC of a term is the set of gates that can influence its expectation value, and can be computed, for instance, using the depth-first search \cite{cormen2009introduction} on the directed acyclic graph representation of the original ansatz. 

\begin{figure}[H]
\centering
\hspace{0.5cm}
\includegraphics[scale=0.125]{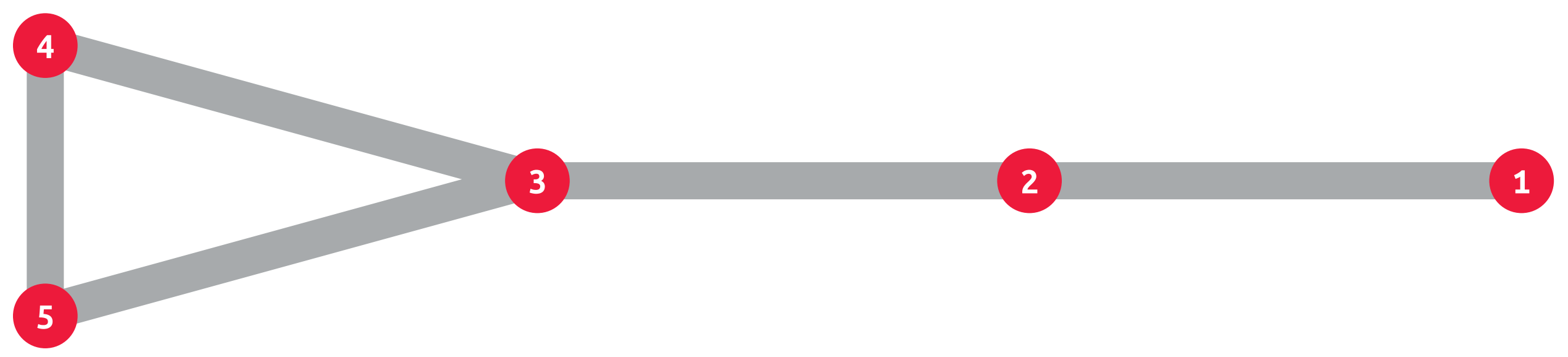}
\caption{The dragon graph, $T_{3,2}$.} \label{fig:dr-graph}
\end{figure}

\begin{figure}[H]
    \begin{subfigure}[b]{0.5\textwidth}
\hspace{0.6cm}\Qcircuit @C=1em @R=.7em {
 \lstick{|0\rangle}&  &  \gate{H}  &  \ctrl{1}  &  \qw  &  \ctrl{1}  &  \qw  &  \qw  &  \qw  &  \gate{X_\beta}  & \qw \\ 
\lstick{|0\rangle} &  &  \gate{H}  &  \targ  &  \gate{Z_{\frac{\gamma}{2}}}  &  \targ  &  \ctrl{1}  &  \qw  &  \ctrl{1}  &  \gate{X_\beta}  & \qw \\ 
\lstick{|0\rangle} &  &  \gate{H}  &  \qw  &  \qw  &  \qw  &  \targ  &  \gate{Z_{\frac{\gamma}{2}}}  &  \targ  &  \qw  & \qw 
}
\label{figsub:z1z2}
        \caption{$Z_1 Z_2$ }
    \end{subfigure}%
    ~ 
    \begin{subfigure}[b]{0.5\textwidth}
        \centering
\scalebox{0.65}{\Qcircuit @C=1em @R=.7em {
\lstick{|0\rangle} &  &  \gate{H}  &  \ctrl{1}  &  \qw  &  \ctrl{1}  &  \qw  &  \qw  &  \qw  &  \qw  &  \qw  &  \qw  &  \qw  &  \qw  &  \qw  &  \qw  & \qw \\ 
\lstick{|0\rangle} &  &  \gate{H}  &  \targ  &  \gate{Z_{\frac{\gamma}{2}}}  &  \targ  &  \ctrl{1}  &  \qw  &  \ctrl{1}  &  \qw  &  \qw  &  \qw  &  \qw  &  \qw  &  \qw  &  \gate{X_\beta}  & \qw \\ 
\lstick{|0\rangle} &  &  \gate{H}  &  \qw  &  \qw  &  \qw  &  \targ  &  \gate{Z_{\frac{\gamma}{2}}}  &  \targ  &  \ctrl{1}  &  \qw  &  \ctrl{1}  &  \ctrl{2}  &  \qw  &  \ctrl{2}  &  \gate{X_\beta}  & \qw \\ 
\lstick{|0\rangle} &  &  \gate{H}  &  \qw  &  \qw  &  \qw  &  \qw  &  \qw  &  \qw  &  \targ  &  \gate{Z_{\frac{\gamma}{2}}}  &  \targ  &  \qw  &  \qw  &  \qw  &  \qw  & \qw \\ 
\lstick{|0\rangle} &  &  \gate{H}  &  \qw  &  \qw  &  \qw  &  \qw  &  \qw  &  \qw  &  \qw  &  \qw  &  \qw  &  \targ  &  \gate{Z_{\frac{\gamma}{2}}}  &  \targ  &  \qw  & \qw 
}}
        \caption{$Z_2 Z_3$}
    \end{subfigure}
    ~
        \begin{subfigure}[b]{0.5\textwidth}
\hspace{0.6cm}\scalebox{0.6}{\Qcircuit @C=1em @R=.7em {
 \lstick{|0\rangle}&  &  \gate{H}  &  \ctrl{1}  &  \qw  &  \ctrl{1}  &  \qw  &  \qw  &  \qw  &  \qw  &  \qw  &  \qw  &  \qw  &  \qw  &  \qw  &  \qw  & \qw \\ 
\lstick{|0\rangle} &  &  \gate{H}  &  \targ  &  \gate{Z_{\frac{\gamma}{2}}}  &  \targ  &  \ctrl{1}  &  \qw  &  \ctrl{1}  &  \qw  &  \qw  &  \qw  &  \ctrl{2}  &  \qw  &  \ctrl{2}  &  \gate{X_\beta}  & \qw \\ 
\lstick{|0\rangle} &  &  \gate{H}  &  \qw  &  \qw  &  \qw  &  \targ  &  \gate{Z_{\frac{\gamma}{2}}}  &  \targ  &  \ctrl{1}  &  \qw  &  \ctrl{1}  &  \qw  &  \qw  &  \qw  &  \gate{X_\beta}  & \qw \\ 
\lstick{|0\rangle} &  &  \gate{H}  &  \qw  &  \qw  &  \qw  &  \qw  &  \qw  &  \qw  &  \targ  &  \gate{Z_{\frac{\gamma}{2}}}  &  \targ  &  \targ  &  \gate{Z_{\frac{\gamma}{2}}}  &  \targ  &  \qw  & \qw 
}}
        \caption{$Z_3 Z_4$}
    \end{subfigure}%
    ~
    \begin{subfigure}[b]{0.5\textwidth}
        \centering
\scalebox{0.75}{\Qcircuit @C=1em @R=.7em {
\lstick{|0\rangle} &  &  \gate{H}  &  \ctrl{1}  &  \qw  &  \ctrl{1}  &  \qw  &  \qw  &  \qw  &  \ctrl{2}  &  \qw  &  \ctrl{2}  &  \qw  & \qw \\ 
\lstick{|0\rangle} &  &  \gate{H}  &  \targ  &  \gate{Z_{\frac{\gamma}{2}}}  &  \targ  &  \ctrl{1}  &  \qw  &  \ctrl{1}  &  \qw  &  \qw  &  \qw  &  \gate{X_\beta}  & \qw \\ 
\lstick{|0\rangle} &  &  \gate{H}  &  \qw  &  \qw  &  \qw  &  \targ  &  \gate{Z_{\frac{\gamma}{2}}}  &  \targ  &  \targ  &  \gate{Z_{\frac{\gamma}{2}}}  &  \targ  &  \gate{X_\beta}  & \qw 
}}
        \caption{$Z_4 Z_5$}
    \end{subfigure}
    ~
        \begin{subfigure}[b]{0.5\textwidth}
\hspace{0.6cm}\Qcircuit @C=1em @R=.7em {
 \lstick{|0\rangle}&  &  \gate{H}  &  \ctrl{1}  &  \qw  &  \ctrl{1}  &  \qw  &  \qw  &  \qw  &  \qw  &  \qw  &  \qw  &  \qw  &  \qw  &  \qw  &  \qw  & \qw \\ 
 \lstick{|0\rangle}&  &  \gate{H}  &  \targ  &  \gate{Z_{\frac{\gamma}{2}}}  &  \targ  &  \ctrl{1}  &  \qw  &  \ctrl{1}  &  \qw  &  \qw  &  \qw  &  \ctrl{2}  &  \qw  &  \ctrl{2}  &  \gate{X_\beta}  & \qw \\ 
\lstick{|0\rangle} &  &  \gate{H}  &  \qw  &  \qw  &  \qw  &  \targ  &  \gate{Z_{\frac{\gamma}{2}}}  &  \targ  &  \ctrl{1}  &  \qw  &  \ctrl{1}  &  \qw  &  \qw  &  \qw  &  \qw  & \qw \\ 
\lstick{|0\rangle} &  &  \gate{H}  &  \qw  &  \qw  &  \qw  &  \qw  &  \qw  &  \qw  &  \targ  &  \gate{Z_{\frac{\gamma}{2}}}  &  \targ  &  \targ  &  \gate{Z_{\frac{\gamma}{2}}}  &  \targ  &  \gate{X_\beta}  & \qw 
}
        \caption{$Z_3 Z_5$}
    \end{subfigure}%

    \caption{Reduced ansatz for each term of the dragon graph Hamiltonian.}
    \label{fig:nra-dg}
\end{figure}
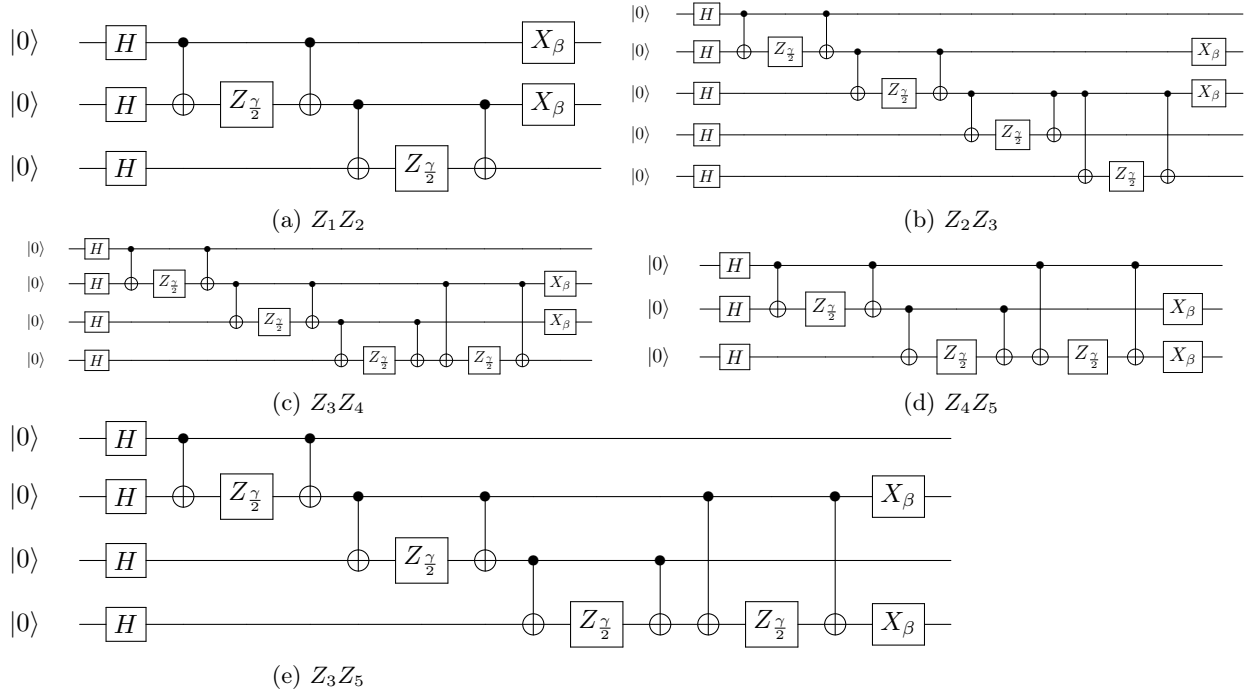

Consider the QAOA ansatz to compute the MAXCUT  of the dragon graph $T_{3,2}$ shown in Figure ~\ref{fig:dr-graph}. One can easily show that the exact MAXCUT for this graph is $4$. The negated QAOA Hamiltonian is $-\frac{1}{2}(5 - Z_1 Z_2 - Z_2 Z_3 - Z_3 Z_4 - Z_4 Z_5 - Z_3 Z_5)$, reflecting the connectivity of the graph. The QAOA ansatz at $p=1$ is shown in Figure ~\ref{fig:banner-ansatz}.  Figure ~\ref{fig:nra-dg} shows the reduced ansatz set for the five observables.

\begin{figure}[H]
\centering
\hspace{0.1cm}\scalebox{0.9}{\Qcircuit @C=1em @R=.7em {
\lstick{|0\rangle} &  &  \gate{H}  &  \ctrl{1}  &  \qw  &  \ctrl{1}  &  \qw  &  \qw  &  \qw  &  \qw  &  \qw  &  \qw  &  \qw  &  \qw  &  \qw  &  \qw  &  \qw  &  \qw  &  \gate{X_\beta}  & \qw \\ 
\lstick{|0\rangle} &  &  \gate{H}  &  \targ  &  \gate{Z_{\frac{\gamma}{2}}}  &  \targ  &  \ctrl{1}  &  \qw  &  \ctrl{1}  &  \qw  &  \qw  &  \qw  &  \qw  &  \qw  &  \qw  &  \qw  &  \qw  &  \qw  &  \gate{X_\beta}  & \qw \\ 
\lstick{|0\rangle} &  &  \gate{H}  &  \qw  &  \qw  &  \qw  &  \targ  &  \gate{Z_{\frac{\gamma}{2}}}  &  \targ  &  \ctrl{1}  &  \qw  &  \ctrl{1}  &  \qw  &  \qw  &  \qw  &  \ctrl{2}  &  \qw  &  \ctrl{2}  &  \gate{X_\beta}  & \qw \\ 
\lstick{|0\rangle} &  &  \gate{H}  &  \qw  &  \qw  &  \qw  &  \qw  &  \qw  &  \qw  &  \targ  &  \gate{Z_{\frac{\gamma}{2}}}  &  \targ  &  \ctrl{1}  &  \qw  &  \ctrl{1}  &  \qw &  \qw  &  \qw  &  \gate{X_\beta}   & \qw \\ 
\lstick{|0\rangle} &  &  \gate{H}  &  \qw  &  \qw  &  \qw  &  \qw  &  \qw  &  \qw  &  \qw  &  \qw  &  \qw  &  \targ  &  \gate{Z_{\frac{\gamma}{2}}}  &  \targ  &  \targ  &  \gate{Z_{\frac{\gamma}{2}}}  &  \targ  &  \gate{X_\beta}  & \qw \\ 
}}
\caption{QAOA ansatz to compute MAXCUT of the dragon $T_{3,2}$ graph at $p = 1$. The ZZ interactions are decomposed using two CNOT and one rotation gates.} 
\label{fig:banner-ansatz}
\end{figure}
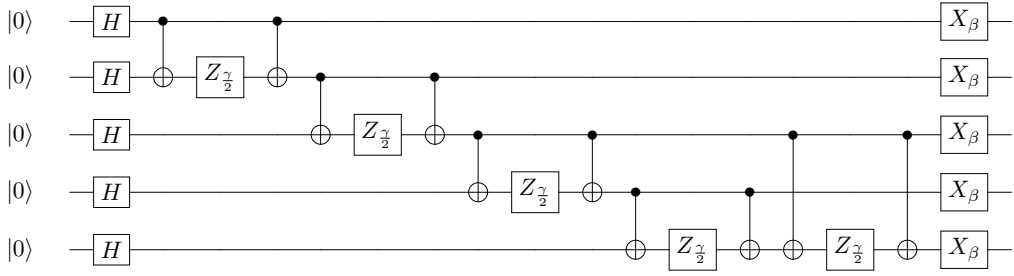

One can compute the expectation value of each term from its reduced ansatz and combine the results to determine the expectation value of the full Hamiltonian. This has a number of advantages over the original VQE/QAOA algorithm. Each reduced ansatz has lower or equal complexity and qubit number compared to the original ansatz. This reduction in both depth and width generally leads to lower noise. We note that for certain problems, there exist special ansatz constructions using a similar approach that possess inherent robustness to noise ~\cite{kim2017robust}, while the current work is a more general method to minimize the effect of noise.  Additionally, some of the sub-Hamiltonians may have lower maximum absolute coefficients which give a tighter upper bound on the variance of the expectation value of the Hamiltonian, or conversely a tighter upper bound on the number of measurements needed to maintain a given variance.  
We experimentally demonstrate this advantage by comparing the energy obtained from the RA-VQE to VQE, and that of RA-QAOA to QAOA. 
Given a qubit Hamiltonian of a VQE problem expressed as $H = \sum^t_\alpha h_\alpha P_\alpha$, has $t$ Pauli observables, one may create $t$ reduced circuits and compute the expectation value for each individually. While giving the most accurate result, this strategy also increases the total number of measurements. 
Instead, once the past causal cone circuits are generated, one can consider the minimum number of reduced ansatz circuits that support all terms in the Hamiltonian, reducing the total number of measurements and hence the time-to-solution. 
For instance, the reduced ansatz for two Hamiltonian terms may share the same circuit when the corresponding Hamiltonian terms are supported by the same qubits but measured in different bases. Similarly, one reduced ansatz may be the subcircuit of another. 

We compute the expectation value for all sub-Hamiltonians independently but minimize them together (or maximize in the case of QAOA) 
with the prescribed number of measurements based on the chosen strategy.  The RA-VQE algorithm can be outlined very coarsely as follows.

\begin{enumerate}
   \item Construct the reduced ansatz set.
   \item Group sub-Hamiltonians based on chosen strategy.
    \item Execute the circuits and measure the expectation value of each observable.
    \item Calculate the expectation value for the overall Hamiltonian.
    \item Use a classical non-linear optimizer to minimize/maximize this expectation value.
\end{enumerate}

\section{Experimental demonstration}
\begin{figure}[H]
\centering
\hspace{0.5cm}
\Qcircuit @C=1em @R=.7em {
\lstick{|0\rangle} & \gate{X_{\pi}} &\targ&\qw &\qw&\qw&\qw& \qw& \qw\\
\lstick{|0\rangle} & \gate{Y_{\phi}} &\ctrl{-1}&\ctrl{1}&\targ&\qw& \qw& \qw& \qw\\
\lstick{|0\rangle} & \qw &\qw&\gate{Y_{\lambda_1}}&\ctrl{-1}&\ctrl{1}&\qw&\targ& \qw\\
\lstick{|0\rangle} &\qw&\qw&\qw&\qw&\gate{Y_{\lambda_2}}&\qw&\ctrl{-1}& \qw
}
\caption{The canonical four qubit UCC ansatz for deuteron \cite{dumitrescu2018cloud}. Here the canonical gate set means the rotation gates, their controlled versions, Hadamard and  CNOT gates.} \label{fig:deu-ansatz}
\end{figure}
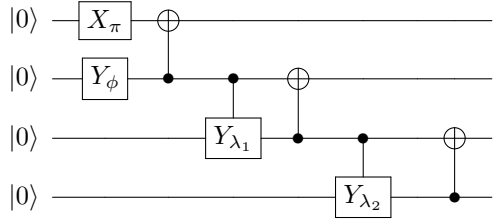

We use the VQE algorithm to compute the binding energy of the deuteron using a pion-less effective field theory. This problem has attracted attention as a benchmark algorithm, and was implemented on both superconducting and a trapped-ion platforms \cite{dumitrescu2018cloud,shehab2019toward}. For a four qubit ansatz, the qubit Hamiltonian is $ 28.657 -2.143 X_0 X_1 -3.913  X_1 X_2  -5.671 X_2  X_3  -2.143 Y_0  Y_1 -3.913 Y_1 Y_2  -5.671 Y_2 Y_3 + 0.218 Z_0 -6.125 Z_1  -9.625 Z_2  -13.125 Z_3$ and the circuit is given in Figure ~\ref{fig:deu-ansatz}.

We use a trapped ion quantum computer to find the deuteron binding energy using both the original and reduced ansatz circuits. Since, the main advantage of our approach is higher process fidelity, we focus on the in-silico global minimum of the energy landscape of $-2.14$ MeV at the parameter values $0.858$, $0.958$, and $0.758$ radians. 
With $5000$ measurements per circuit, we determine the experimental binding energy as $-1.5(2)$ MeV for the original VQE ansatz, and $-2.0(2)$ MeV for the reduced ansatz. This is a $\sim 80\%$ improvement in accuracy, making the energy consistent with the theoretical value. Grouping of the sub-Hamiltonians gives a similar result.
When we prioritize time-to-solution, $14600$ measurements are sufficient to determine the binding energy as $-1.5(3)$ MeV. This is a $\sim 70\%$ reduction compared to the $50000$ measurements the original VQE ansatz needed. 

\begin{figure}[H]
    \begin{subfigure}[b]{0.5\textwidth}
\includegraphics[scale=0.4]{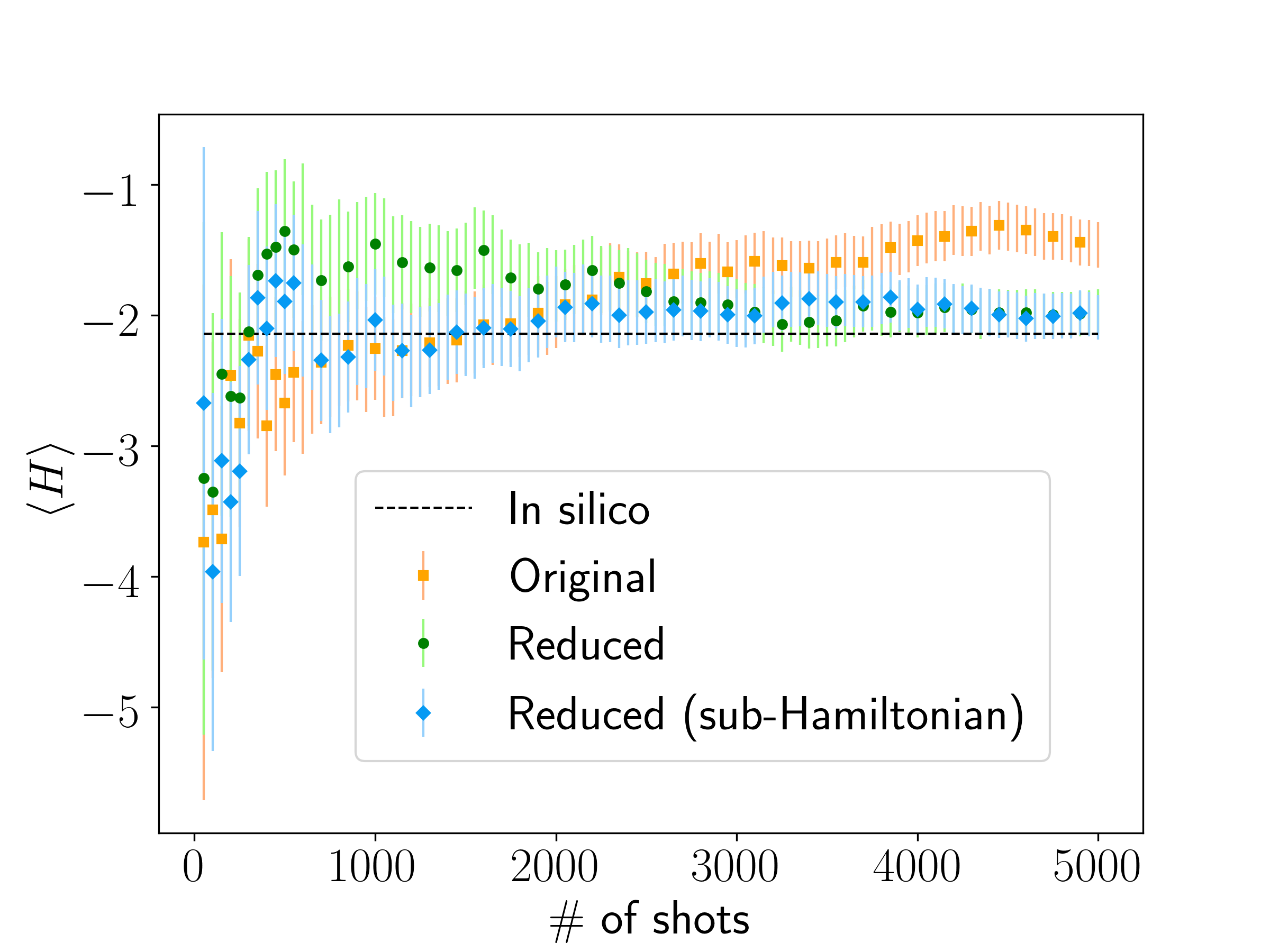}
        \caption{Binding energy vs. \# of measurements}
    \end{subfigure}%
    ~ 
    \begin{subfigure}[b]{0.5\textwidth}
        \centering
\includegraphics[scale=0.4]{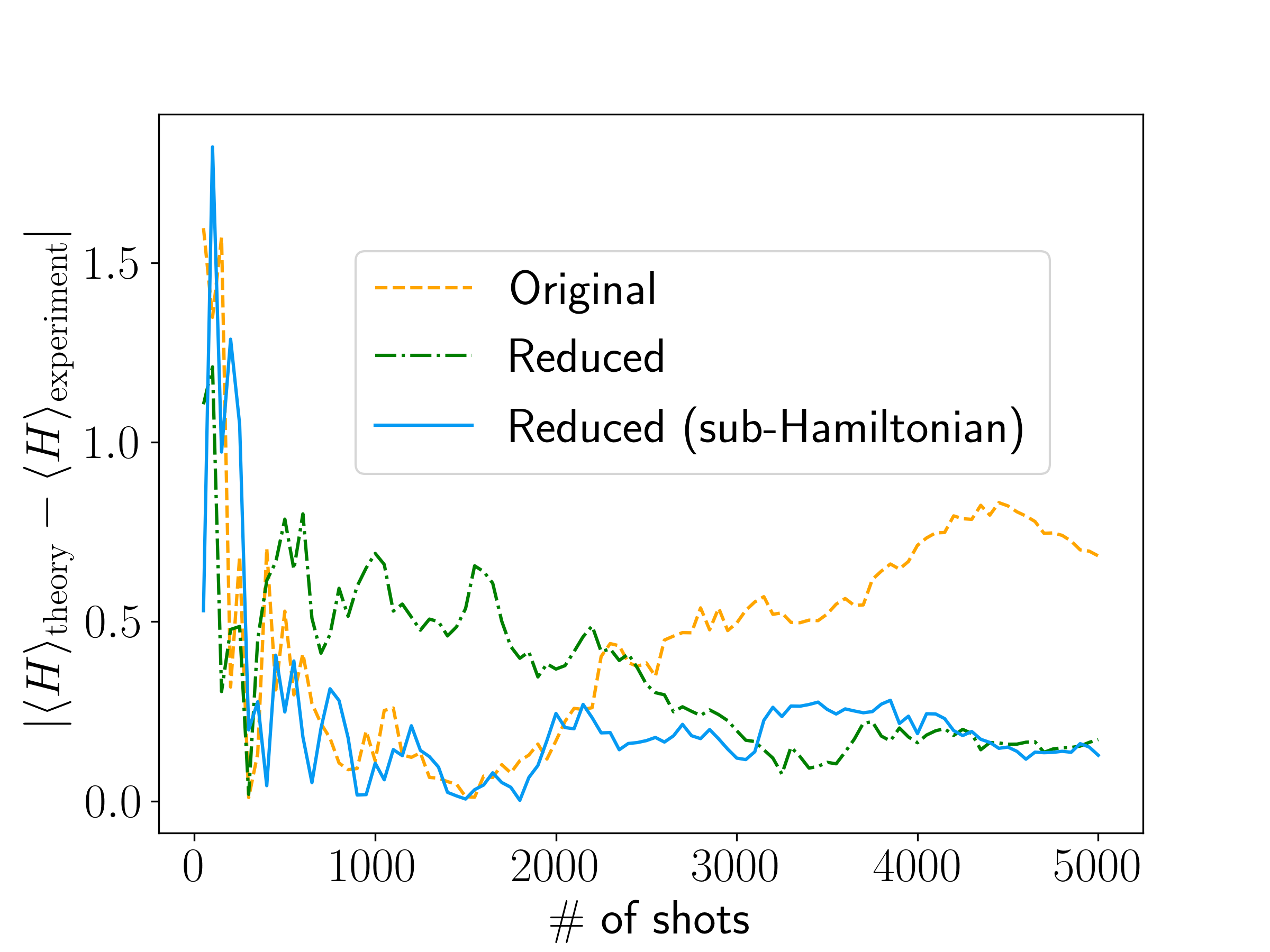}
\caption{Absolute standard error vs. \# of measurements}
    \end{subfigure}
\caption{Theoretical and experimentally determined binding energies (a) and the absolute standard errors (b) for the original and the reduced VQE ansatz. Data for all individual Hamiltonian terms is reported in Appendix ~\ref{sec:termwiseplot}.}
    \label{fig:result}
\end{figure}

The experimentally determined binding energies and the absolute standard errors are plotted against increasing number of measurements per circuit in Figure ~\ref{fig:result}. Appendix ~\ref{sec:termwiseplot} contains similar results for each individual Hamiltonian term.

We also demonstrate the advantage of the RA-QAOA ansatz on the  trapped ion quantum computer by solving the MAXCUT problem for the dragon graph introduced in Section ~\ref{sec:nra}. We again use the set of parameters corresponding to the in-silico global minimum of $-3.45$, which are $\gamma = 1.358 $, and $\beta = 2.462$. With $5000$ measurements per circuit, the original QAOA ansatz gives $-3.26(2)$ while the reduced version gives $-3.33(2)$, a $\sim 36(9)\%$ improvement in accuracy. 

 $5410$ measurements ($5500$ measurements conducted in the experiment) are sufficient to determine the MAXCUT as $-3.34(7)$ . This is a $\sim 78\%$ reduction in the number of measurements compared to $25000$ measurements needed in the original QAOA ansatz. 

\begin{figure}[H]
    \begin{subfigure}[b]{0.5\textwidth}
\includegraphics[scale=0.4]{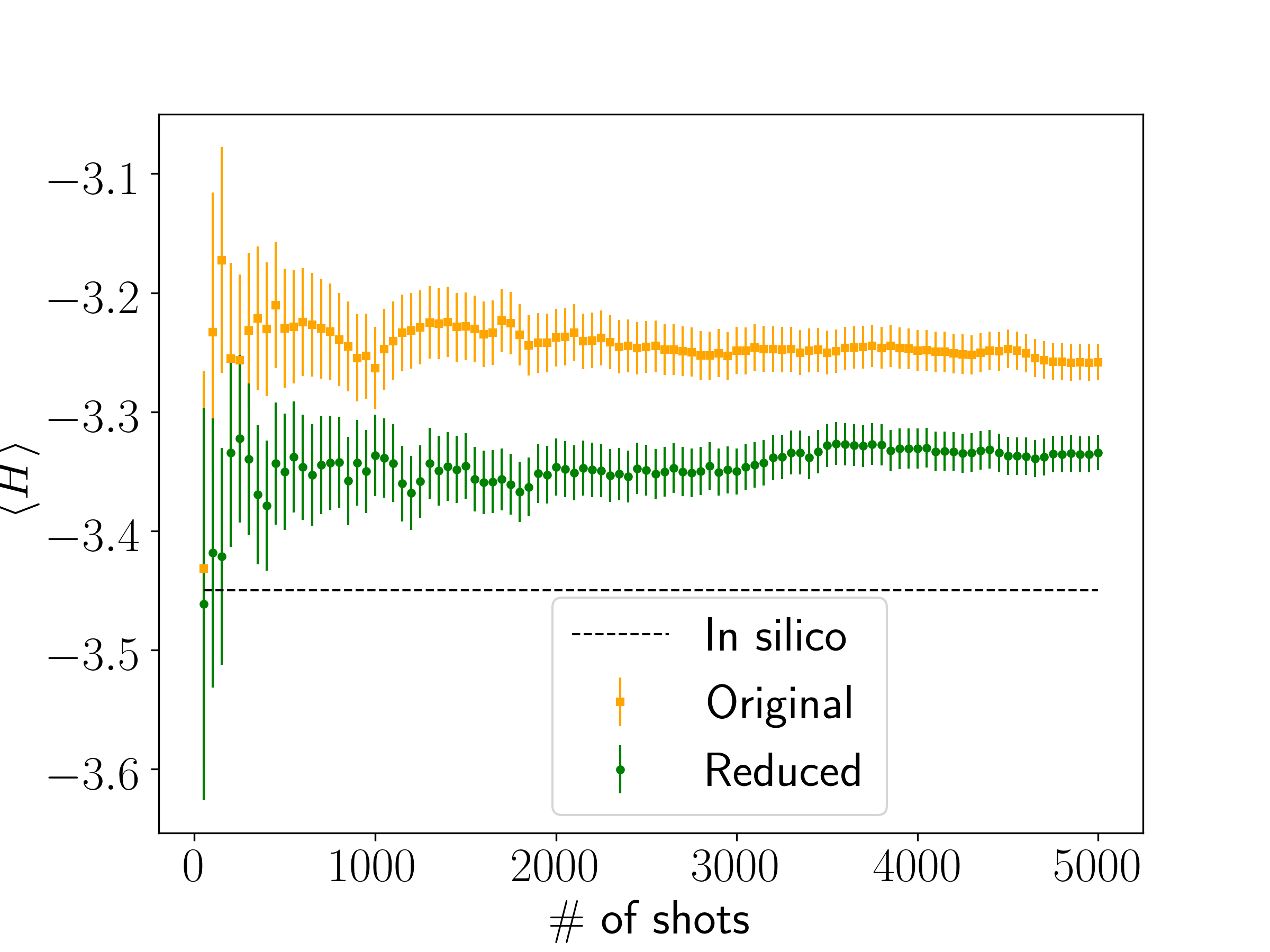}
        \caption{MAXCUT vs. \# of measurements}
    \end{subfigure}%
    ~ 
    \begin{subfigure}[b]{0.5\textwidth}
        \centering
\includegraphics[scale=0.4]{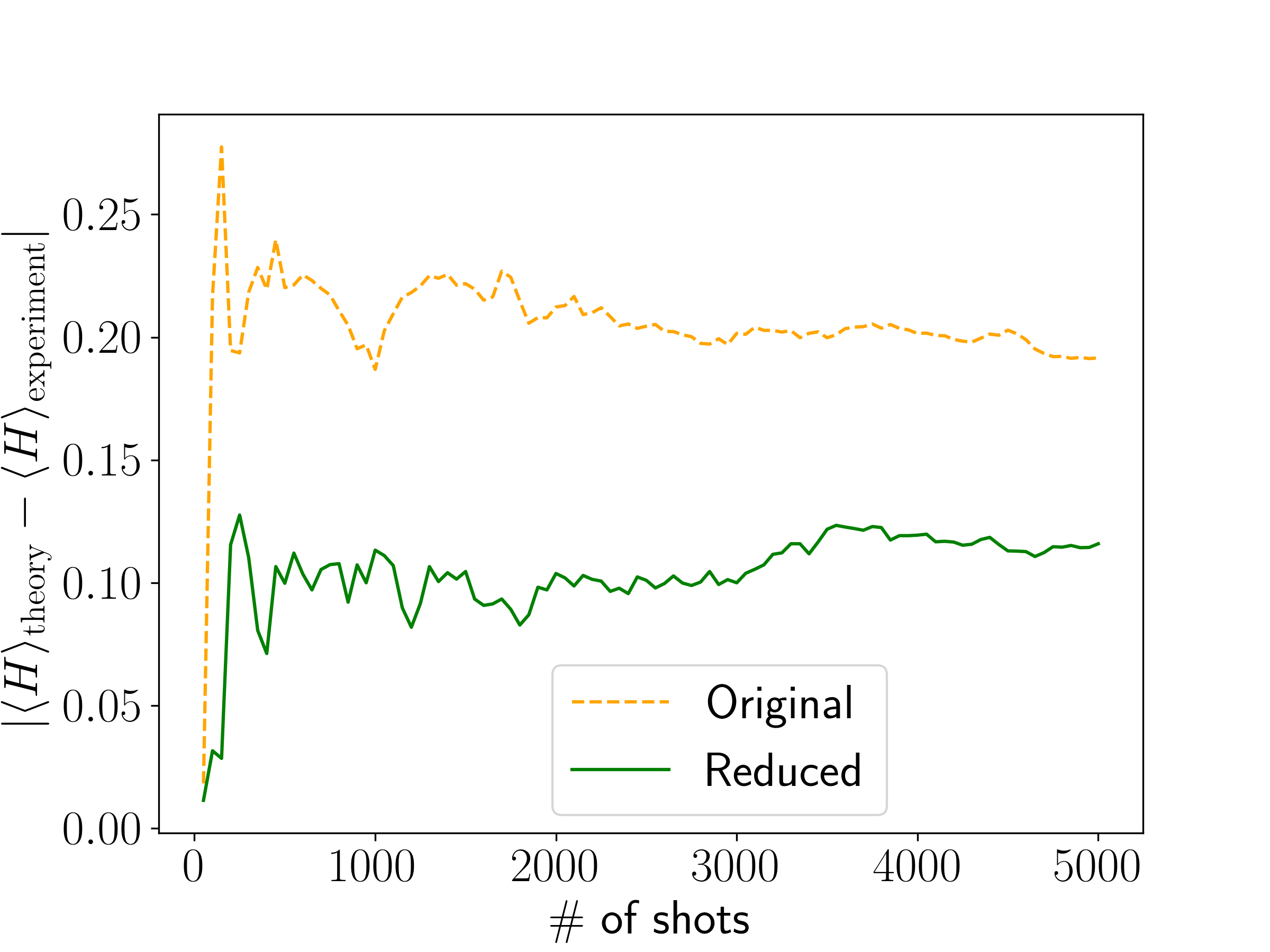}
\caption{Absolute standard error vs. \# of measurements}
    \end{subfigure}
\caption{Theoretical and experimentally determined MAXCUT (a) and the absolute standard errors (b) for the original QAOA ansatz and the reduced ansatz at the parameters ($\gamma = 1.358, \beta = 2.462, p = 1$). Similar experimental data for individual Hamiltonian term is reported in Appendix ~\ref{sec:termwiseplot}.}
    \label{fig:result-dragon}
\end{figure}

The experimentally determined MAXCUT and the absolute standard errors are plotted against increasing number of measurements per circuit in Figure ~\ref{fig:result-dragon}. Appendix ~\ref{sec:termwiseplot} contains similar results for each individual Hamiltonian terms. For both examples, our method achieves more accurate results with fewer measurements than the standard VQE/QAOA circuits.

\section{Discussion}
\label{sec:disc}
The reduced ansatz methods developed and demonstrated here show how targeted circuit optimization can give significant performance increases which are crucial for NISQ devices. Depending on the problem, the design of the original ansatz can even be informed by its potential to take advantage of subsequent reduced ansatz formulations. In the future, we hope to investigate  how the structural complexity of a problem may adversely affect the advantages expected to be achieved from the reduced ansatz approach, and how this method can be adapted to other types of algorithms.

\section{Methods}

\subsection{Trapped ion hardware}
The trapped-ion quantum computer uses the $^{2}S_{1/2} |F=0,m_F=0\rangle $ and $|F=1,m_F =1\rangle $ states of individual trapped $^{171}\text{Yb}^{+}$ ions as qubits. The ions are initialized by optical pumping to $ |F=0,m_F=0\rangle $ and detected by state-dependent fluorescence on the $2S_{\frac{1}{2}}$ to $2P_{\frac{1}{2}}$ transition. We use a pair of counter-propagating Raman beams, one of which is split into an array of individual addressing beams, to drive gate operations. 
The two native gates in the system are single-qubit R-gates which are rotations around an axis in the X/Y plane, single qubit Z-rotations by phase advances in the classical controllers, and two qubit entangling XX-gates which use the motional modes to create entanglement between any two qubits. Both the R as well as the XX angle can be varied continuously. For the details of the single and two qubit gate implementations we refer the reader to Appendix A of \cite{landsman2018verified} and to \cite{choi2014optimal, debnath2016demonstration, molmer1999multiparticle, zhu2006arbitrary}. Typical gate times are $10 \mu s$ for single-qubit and $210 \mu s$ for XX-gates. The errors in state initialization and detection are corrected by applying the inverse of an independently measured state-to-state error matrix. Typical gate fidelities are $\approx 99.5\%$ for single qubit gates and $\approx 98.5\%$ for XX-gates.

For the four-qubit deuteron ansatz, seven ions are loaded into the trap, where the inner five are used as qubits, with the outermost pair being used to evenly space the middle five ions. The algorithmic qubits $1,2,3,4$ are mapped onto physical qubits $1,2,3,5$. The average four-qubit readout fidelity is $97.1\%$. To be consistent, the same physical qubits are used for the reduced ansatz.

For the dragon graph ansatz, the algorithmic qubits $1,2,3,4, 5$ are mapped onto physical qubits $1,5,3,2,4$ 
. The average read-out fidelity for five qubits is $94.3 \%$. Similarly, the same mapping is used for the reduced ansatz. 

Error bars for the correlators are the one-sigma intervals of the asymmetric binomial distribution of state populations. Since the error bars for the correlators tend to a symmetric limit for large number of shots, the error bar for the Hamiltonian can be approximated by a Gaussian distribution, which follows from propagation of the errors of individual correlators.

\subsection{Construction of the reduced ansatz for the deuteron}
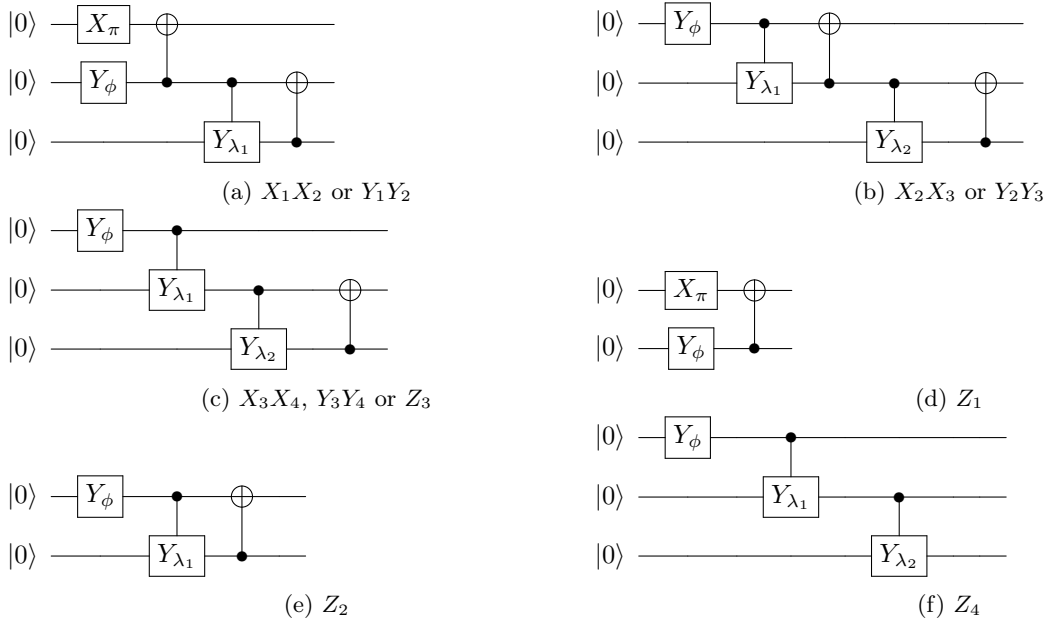
\begin{figure}[H]
    \begin{subfigure}[b]{0.5\textwidth}
\hspace{0.6cm}\Qcircuit @C=1em @R=.7em {
\lstick{|0\rangle} & \gate{X_{\pi}} &\targ&\qw &\qw&\qw\\
\lstick{|0\rangle} & \gate{Y_{\phi}} &\ctrl{-1}&\ctrl{1}&\targ&\qw\\
\lstick{|0\rangle} & \qw &\qw&\gate{Y_{\lambda_1}}&\ctrl{-1}&\qw
}
\label{figsub:x1x2}
        \caption{$X_1 X_2$ or $Y_1 Y_2$}
    \end{subfigure}%
    ~ 
    \begin{subfigure}[b]{0.5\textwidth}
        \centering
\Qcircuit @C=1em @R=.7em {
\lstick{|0\rangle} & \gate{Y_{\phi}} &\ctrl{1}&\targ&\qw& \qw& \qw& \qw\\
\lstick{|0\rangle} & \qw&\gate{Y_{\lambda_1}}&\ctrl{-1}&\ctrl{1}&\qw&\targ& \qw\\
\lstick{|0\rangle} &\qw&\qw&\qw&\gate{Y_{\lambda_2}}&\qw&\ctrl{-1}& \qw
}
        \caption{$X_2 X_3$ or $Y_2 Y_3$}
    \end{subfigure}
    ~
        \begin{subfigure}[b]{0.5\textwidth}
\hspace{0.6cm}\Qcircuit @C=1em @R=.7em {
\lstick{|0\rangle} & \gate{Y_{\phi}} &\ctrl{1}&\qw& \qw& \qw& \qw\\
\lstick{|0\rangle}  &\qw&\gate{Y_{\lambda_1}}&\ctrl{1}&\qw&\targ& \qw\\
\lstick{|0\rangle} &\qw&\qw&\gate{Y_{\lambda_2}}&\qw&\ctrl{-1}& \qw
}
        \caption{$X_3 X_4$, $Y_3 Y_4$ or $Z_3$}
    \end{subfigure}%
    ~
    \begin{subfigure}[b]{0.5\textwidth}
        \centering
\Qcircuit @C=1em @R=.7em {
\lstick{|0\rangle} & \gate{X_{\pi}} &\targ& \qw\\
\lstick{|0\rangle} & \gate{Y_{\phi}} &\ctrl{-1}& \qw
}
        \caption{$Z_1$}
    \end{subfigure}
    ~
        \begin{subfigure}[b]{0.5\textwidth}
\hspace{0.6cm}\Qcircuit @C=1em @R=.7em {
\lstick{|0\rangle} & \gate{Y_{\phi}} &\ctrl{1}&\targ&\qw& \qw\\
\lstick{|0\rangle}  &\qw&\gate{Y_{\lambda_1}}&\ctrl{-1}&\qw& \qw
}
        \caption{$Z_2$}
    \end{subfigure}%
    ~
        \begin{subfigure}[b]{0.5\textwidth}
        \centering
\Qcircuit @C=1em @R=.7em {
\lstick{|0\rangle} & \gate{Y_{\phi}} &\qw&\ctrl{1}&\qw&\qw& \qw& \qw& \qw\\
\lstick{|0\rangle} & \qw &\qw&\gate{Y_{\lambda_1}}&\qw&\ctrl{1}&\qw&\qw& \qw\\
\lstick{|0\rangle} &\qw&\qw&\qw&\qw&\gate{Y_{\lambda_2}}&\qw&\qw& \qw
}
        \caption{$Z_4$}
    \end{subfigure}
    \caption{Reduced ansatz for each term of the deuteron Hamiltonian.}
    \label{fig:nra-deut}
\end{figure}

In our approach, the original VQE ansatz is divided into smaller ansatz circuits, one for each term in the Hamiltonian as shown in  Figure ~\ref{fig:nra-deut}.

\subsection{Construction of the deuteron sub-Hamiltonians}
Table ~\ref{tab:sh-nr-d} shows the Hamiltonian terms supported by every reduced ansatz of the deuteron problem.

\begin{table}[H]
\centering
\begin{tabular}{|c |  p{11cm}|}
\hline
{\bf Term} &  {\bf Sub-Hamiltonian}\\
\hline
\hline
$X_1 X_2$ or $Y_1 Y_2$   & $ 28.657  I -2.143  X_1 X_2 
 -2.143 Y_1 Y_2  + 0.218 Z_1 -6.125 Z_2 
 $\\
 \hline
$X_2 X_3$ or $Y_2 Y_3$  & $ 28.657  I -3.913  X_2 X_3 -5.671 X_3 X_4
 -3.913 Y_2 Y_3 -5.671 Y_3 Y_4  -6.125 Z_2 
 -9.625 Z_3 -13.125 Z_4$\\
\hline
$X_3 X_4$ or $Y_3 Y_4$ or $Z_3$  & $ 28.657  I -5.671 X_3 X_4
 -5.671 Y_3 Y_4  
 -9.625 Z_3 -13.125 Z_4$\\
\hline
$Z_1$  &$ 28.657  I   + 0.218 Z_1  
$\\
\hline
$Z_2$   &$ 28.657  I    -6.125 Z_2 
 $\\
\hline
$Z_4$   &$ 28.657  I  -13.125 Z_4$\\
\hline
\end{tabular}
\caption{New sub-Hamiltonian simulation problems generated by the reduced approach}
\label{tab:sh-nr-d}
\end{table}

One can either run all the ansatz circuits in Table ~\ref{tab:sh-nr-d} to prioritize accuracy, or a minimal subset which covers every term to reduce time-to-solution. We experimentally demonstrate that both strategies with the same number of measurements as in the original VQE algorithm, and determine the binding energy  more accurately than the  the original VQE ansatz. For the second strategy, we  consider the first two sub-Hamiltonians. $\langle Z_2 \rangle$ is considered as a term of only the first sub-Hamiltonian to avoid repeated calculation since the corresponding ansatz is shallower.

\subsection{Estimating the number of measurements for shorter time-to-solution}
If accuracy is the priority, one should run the reduced VQE or QAOA ansatz with as many measurements as possible. We run $5000$ measurements per circuit, which would require  $50000$ measurements  in total for ten reduced ansatz circuits.  If the accuracy of the original VQE or QAOA ansatz is sufficient, it may be achieved with fewer  measurements with the reduced ansatz. 

Since, both the original and reduced ansatz would not be used together in practice, the target accuracy can only be estimated from the previous experiments of the same scale. We run the original VQE ansatz and use a standard 1-$\sigma$ 
error to estimate the number of measurements needed for the reduced approach to maintain the same error. The target error rate, the absolute value of the coefficient of the target observable ($h_\gamma$), and the coefficient of largest absolute value ($h_{\gamma, max}$) are used to estimate the number of measurements $S_\beta$ according to $\epsilon = \sqrt{\frac{T|h^2_{\text{max}}|}{S}}$ which is
equation 12 of the supplemental material of \cite{kandala2017hardware}. The results are given in table  Table ~\ref{tab:dc-nr-meta}. In this table, $h_\gamma$ is the coefficient of the corresponding Hamiltonian term, $h_{\gamma, max}$ is the coefficient with the largest absolute value in the sub-Hamiltonian, and $T_\beta$ is the number of terms in the corresponding sub-Hamiltonian. In the experiment we use the closest multiple of fifty as the prescribed number of measurements. When the prescribed number is too small we replace it with $~500$ to avoid the initial  fluctuation.

\begin{table}[H]
\centering
\begin{tabular}{|c | p{3cm}| c|p{2cm}|p{2cm}|}
\hline
{\bf Term}  & $|h_\gamma|$  & $|h_{\gamma, max}|$ & $\text{\bf T}_\beta$ & $\text{\bf S}_\beta$\\
\hline
\hline
$X_1 X_2$ or $Y_1 Y_2$   &$2.143$& $6.125$&$4$&$\sim 436$\\
 \hline
$X_2 X_3$ or $Y_2 Y_3$ & $3.913$ &$13.125$&$7$&$\sim 3500$\\
\hline
$X_3 X_4$ or $Y_3 Y_4$ or $Z_3$  & $5.671$ or $9.625$  &$13.125$&$4$&$\sim 2000$\\
\hline
$Z_1$   & $0.218$  &$0.218$&$1$&$\sim 1$\\
\hline
$Z_2$   & $6.125$ & $6.125$&$1$&$\sim 109$\\
\hline
$Z_4$   & $13.125$ & $13.125$&$1$&$\sim 500$\\
\hline
\end{tabular}
\caption{Estimated number of measurements for the deuteron ansatz to maintain the original VQE accuracy. $S_\beta$ is the number of estimated measurements, $h_\gamma$ is the coefficient of the corresponding Hamiltonian term, $h_{\gamma, max}$ is the coefficient with the largest absolute value in the sub-Hamiltonian, and $T_\beta$ is the number of terms in the corresponding sub-Hamiltonian.}
\label{tab:dc-nr-meta}
\end{table}
In a similar manner, the number of measurements per Hamiltonian term needed for the reduced-ansatz QAOA approach to maintain the similar error level as in the original QAOA algorithm is determined as $1082$. In the experiment, we use $1100$  measurements with the goal to maintain the  accuracy $0.034$ found for the original QAOA ansatz.

\subsection{Circuit optimization for trapped ion hardware}
The following circuit  identities  are used to translate the canonical gates into the physical gates native to our trapped ion architecture \cite{maslov2017basic, maslov2018use}.

\begin{align}
\label{eq:iden}
\raisebox{-0.01em}{\Qcircuit @C=1em @R=1.5em {
  &\ctrl{1}& \qw\\
 &\targ&\qw
}} 
&\raisebox{-1.15em}{\hspace{2mm}=\hspace{4mm}}
\scalebox{0.9}{\Qcircuit @C=1em @R=.7em {
 & \gate{Y_{\frac{\pi}{2}}} & \multigate{1}{XX_{\frac{\pi}{2}}}& \gate{X_{-\frac{\pi}{2}}}& \gate{Y_{-\frac{\pi}{2}}}&  \qw\\
 & \qw&\ghost{XX_{\frac{\pi}{2}}}&\gate{X_{-\frac{\pi}{2}}}&\qw&\qw
}}
\nonumber\\
\Qcircuit @C=1em @R=.9em {
 & \ctrl{1}& \qw\\
 & \gate{Y_{\theta}}&\qw
} 
&\raisebox{-1.15em}{\hspace{2mm}=\hspace{4mm}}
\Qcircuit @C=1em @R=.7em {
 & \qw& \ctrl{1}& \qw& \ctrl{1}& \qw\\
 & \gate{Y_{\frac{\theta}{2}}}&\targ&\gate{Y_{-\frac{\theta}{2}}}&\targ&\qw
}
\nonumber\\
\raisebox{-0.01em}{\Qcircuit @C=1em @R=.7em {
 & \ctrl{1}&\qw &\ctrl{1}& \qw\\
 & \targ& \gate{Y_{\theta}}&\targ&\qw
}} 
&\raisebox{-1.15em}{\hspace{2mm}=\hspace{4mm}}
\scalebox{0.8}{\Qcircuit @C=1em @R=.7em {
 & \gate{X_{-\frac{\pi}{2}}}&\gate{Z_{-\frac{\pi}{2}}} & \multigate{1}{XX_{\theta}}&\gate{Z_{\frac{\pi}{2}}}&\gate{X_{\frac{\pi}{2}}}&  \qw\\
 & \qw&\gate{Z_{-\frac{\pi}{2}}}&\ghost{XX_{\theta}}&\gate{Z_{\frac{\pi}{2}}}&\qw
}}
\end{align}

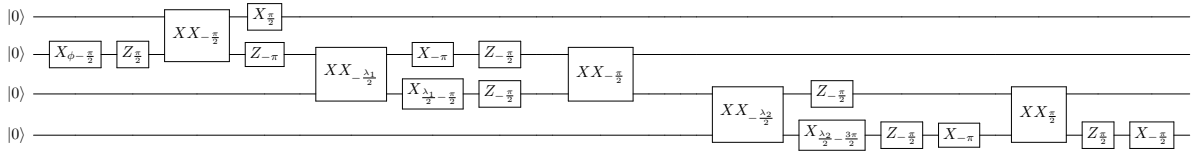
\begin{figure}[H]
\centering
\scalebox{0.6}{
\Qcircuit @C=1em @R=.7em {
\lstick{|0\rangle}  &\qw &\qw&\multigate{1}{XX_{-\frac{\pi}{2}}} &\gate{X_{\frac{\pi}{2}}}&\qw &\qw&\qw&\qw&\qw& \qw&\qw &\qw&\qw&\qw&\qw&\qw& \qw & \qw& \qw& \qw&\qw&\qw&\qw&\qw\\
\lstick{|0\rangle} &\gate{X_{\phi - \frac{\pi}{2}}} &\gate{Z_{\frac{\pi}{2}}}&\ghost{XX_{-\frac{\pi}{2}}}&\gate{Z_{-\pi}}&\qw& \multigate{1}{XX_{-\frac{\lambda_1}{2}}}&\gate{X_{-\pi}}&\gate{Z_{-\frac{\pi}{2}}}&\qw&  \qw&\multigate{1}{XX_{-\frac{\pi}{2}}}&\qw&\qw&\qw& \qw &\qw&\qw&\qw&\qw&\qw&\qw&\qw&\qw&\qw\\
\lstick{|0\rangle}&\qw&\qw&\qw&\qw&\qw&\ghost{XX_{-\frac{\lambda_1}{2}}}&\gate{X_{\frac{\lambda_1}{2}-\frac{\pi}{2}}}&\gate{Z_{-\frac{\pi}{2}}}&\qw&\qw&\ghost{XX_{-\frac{\pi}{2}}}&\qw&\qw&\qw&\qw&\multigate{1}{XX_{-\frac{\lambda_2}{2}}}&\gate{Z_{-\frac{\pi}{2}}}&\qw&\qw&\qw&\multigate{1}{XX_{\frac{\pi}{2}}}&\qw &\qw & \qw\\
\lstick{|0\rangle} &\qw&\qw&\qw&\qw&\qw&\qw&\qw&\qw&\qw&\qw&\qw&\qw&\qw&\qw&\qw&\ghost{XX_{-\frac{\lambda_2}{2}}}&\gate{X_{\frac{\lambda_2}{2}-\frac{3 \pi}{2}}}&\gate{Z_{-\frac{\pi}{2}}}&\gate{X_{-\pi}}&\qw&\ghost{XX_{\frac{\pi}{2}}}&\gate{Z_{\frac{\pi}{2}}}&\gate{X_{-\frac{\pi}{2}}}& \qw 
}}
\caption{Optimized four qubit deuteron ansatz circuit , written over a native gate set for our trapped-ion quantum computer.} \label{fig:opt4}
\end{figure}

These circuits are then optimized using known rules (refer to  \cite{maslov2018use} for details). The goal is to reduce the number of XX  and RX gates. The optimized physical version of the original VQE ansatz to compute the binding energy of deuteron is given in Figure ~\ref{fig:opt4}.

\begin{figure}[H]
    \begin{subfigure}[b]{\textwidth}
\hspace{0.6cm}\scalebox{0.9}{\Qcircuit @C=1em @R=.7em {
\lstick{|0\rangle} & \qw &\qw&\multigate{1}{XX_{-\frac{\pi}{2}}}&\gate{X_{\frac{\pi}{2}}}&\qw&\qw &\qw&\qw&\qw&\qw&\qw&\qw&\qw\\
\lstick{|0\rangle} & \gate{X_{\phi - \frac{\pi}{2}}}&\gate{Z_{\frac{\pi}{2}}} &\ghost{XX_{-\frac{\pi}{2}}}&\gate{Z_{-\pi}}&\qw&\multigate{1}{XX_{-\frac{\lambda_1}{2}}}&\gate{X_{-\pi}}&\gate{Z_{-\frac{\pi}{2}}}&\multigate{1}{XX_{-\frac{\pi}{2}}}&\qw&\qw&\qw&\qw\\
\lstick{|0\rangle} &\qw& \qw &\qw&\qw&\qw&\ghost{XX_{-\frac{\lambda_1}{2}}}&\gate{X_{\frac{\lambda_1}{2}-\frac{\pi}{2}}}&\gate{Z_{-\frac{\pi}{2}}}&\ghost{XX_{-\frac{\pi}{2}}}&\gate{X_{-\frac{\pi}{2}}}&\gate{Z_{-\frac{\pi}{2}}}&\gate{X_{-\frac{\pi}{2}}}&\qw
}}
\label{figsub:x1x2}
        \caption{$X_1 X_2$ or $Y_1 Y_2$}
    \end{subfigure}
    
    \begin{subfigure}[b]{\textwidth}
        \centering
\hspace{0.6cm}\scalebox{0.8}{
\Qcircuit @C=1em @R=.7em {
\lstick{|0\rangle} & \gate{X_{-\frac{\pi}{2}}} &\gate{Z_{-\phi-\frac{\pi}{2}}}&\multigate{1}{XX_{-\frac{\lambda_1}{2}}}&\gate{X_{-\pi}}&\gate{Z_{-\frac{\pi}{2}}}&\multigate{1}{XX_{-\frac{\pi}{2}}}&\qw&\qw&  \qw&\qw&\qw&\qw& \qw& \qw& \qw\\
\lstick{|0\rangle} & \qw&\qw&\ghost{XX_{-\frac{\lambda_1}{2}}}&\gate{X_{\frac{\lambda_1}{2}-\frac{3 \pi}{2}}}&\gate{Z_{\frac{\pi}{2}}}&\ghost{XX_{-\frac{\pi}{2}}}&\qw&\qw&\multigate{1}{XX_{-\frac{\lambda_2}{2}}}&\gate{X_{-\pi}}&\gate{Z_{-\frac{\pi}{2}}}&\multigate{1}{XX_{\frac{\pi}{2}}}&\qw& \qw&\qw\\
\lstick{|0\rangle} &\qw&\qw&\qw&\qw&\qw&\qw&\qw&\qw&\ghost{XX_{-\frac{\lambda_2}{2}}}&\gate{X_{\frac{\lambda_2}{2}-\frac{\pi}{2}}}&\gate{Z_{-\frac{3 \pi}{2}}}&\ghost{XX_{\frac{\pi}{2}}}&\gate{Z_{\frac{\pi}{2}}}& \gate{X_{-\frac{\pi}{2}}}&\qw
}
}
        \caption{$X_2 X_3$ or $Y_2 Y_3$}
    \end{subfigure}
    
        \begin{subfigure}[b]{\textwidth}
\hspace{0.6cm}\scalebox{0.8}{
\Qcircuit @C=1em @R=.7em {
\lstick{|0\rangle} & \gate{X_{-\frac{\pi}{2}}} &\gate{Z_{-\phi-\frac{\pi}{2}}}&\multigate{1}{XX_{-\frac{\lambda_1}{2}}}&\gate{X_{-\pi}}&\gate{Z_{-\frac{\pi}{2}}}&\gate{X_{-\frac{\pi}{2}}}&\qw&\qw& \qw& \qw&\qw&  \qw&\qw&  \qw\\
\lstick{|0\rangle} & \qw&\qw&\ghost{XX_{-\frac{\lambda_1}{2}}}&\gate{X_{\frac{\lambda_1}{2}-\frac{ \pi}{2}}}&\qw&\gate{Z_{-\frac{\pi}{2}}}&\multigate{1}{XX_{-\frac{\lambda_2}{2}}}&\gate{X_{-\frac{\pi}{2}}}&\gate{Z_{-\frac{\pi}{2}}}&\qw&\multigate{1}{XX_{\frac{\pi}{2}}}&\qw&\qw& \qw\\
\lstick{|0\rangle} &\qw&\qw&\qw&\qw&\qw&\qw&\ghost{XX_{-\frac{\lambda_2}{2}}}&\gate{X_{\frac{\lambda_2}{2}-\frac{\pi}{2}}}&\gate{Z_{-\frac{3 \pi}{2}}}&\qw&\ghost{XX_{\frac{\pi}{2}}}&\gate{Z_{\frac{\pi}{2}}}&\gate{X_{-\frac{\pi}{2}}}& \qw
}
}
        \caption{$X_3 X_4$, $Y_3 Y_4$ or $Z_3$}
    \end{subfigure}
    
    \begin{subfigure}[b]{\textwidth}
        \centering
\hspace{0.7cm} \scalebox{1}{
\Qcircuit @C=1em @R=.7em {
\lstick{|0\rangle} & \gate{X_{\pi}}&\qw &\qw&\multigate{1}{XX_{\frac{\pi}{2}}}&\gate{X_{\frac{\pi}{2}}}&\qw& \qw\\
\lstick{|0\rangle} & \gate{X_{\frac{\pi}{2} - \phi}}&\gate{Z_{-\frac{\pi}{2}}} &\qw&\ghost{XX_{\frac{\pi}{2}}}&\gate{Z_{\frac{\pi}{2}}}&\gate{X_{-\frac{\pi}{2}}}& \qw
}
}
        \caption{$Z_1$}
    \end{subfigure}
    
        \begin{subfigure}[b]{\textwidth}
\hspace{0.6cm}\scalebox{1}{
\Qcircuit @C=1em @R=.7em {
\lstick{|0\rangle} & \gate{X_{-\frac{\pi}{2}}} &\gate{Z_{-\phi-\frac{\pi}{2}}}&\multigate{1}{XX_{-\frac{\lambda_1}{2}}}&\gate{X_{-\pi}}&\gate{Z_{-\frac{\pi}{2}}}&\qw&\qw&\multigate{1}{XX_{\frac{\pi}{2}}}&\gate{X_{-\pi}}&\qw& \qw\\
\lstick{|0\rangle}  &\qw&\qw&\ghost{XX_{-\frac{\lambda_1}{2}}}&\qw&\qw&\gate{X_{\frac{\lambda_1}{2} + \frac{\pi}{2}}}&\gate{Z_{\frac{\pi}{2}}}&\ghost{XX_{\frac{\pi}{2}}}&\gate{Z_{-\frac{\pi}{2}}}&\gate{X_{-\frac{\pi}{2}}}& \qw
}
}
        \caption{$Z_2$}
    \end{subfigure}
    
        \begin{subfigure}[b]{\textwidth}
        \centering
\hspace{0.6cm}\scalebox{1}{
\Qcircuit @C=1em @R=.7em {
\lstick{|0\rangle} & \gate{X_{-\frac{\pi}{2}}} &\gate{Z_{-\phi-\frac{\pi}{2}}}&\multigate{1}{XX_{-\frac{\lambda_1}{2}}}&\gate{X_{-\pi}}&\gate{Z_{-\frac{\pi}{2}}}&\gate{X_{-\frac{\pi}{2}}}& \qw& \qw& \qw&\qw&\qw& \qw\\
\lstick{|0\rangle}  &\qw&\qw&\ghost{XX_{-\frac{\lambda_1}{2}}}&\gate{X_{\frac{\lambda_1}{2}-\frac{\pi}{2}}}&\qw&\gate{Z_{-\frac{\pi}{2}}}&\qw&\multigate{1}{XX_{-\frac{\lambda_2}{2}}}&\gate{X_{-\frac{\pi}{2}}}&\gate{Z_{-\frac{\pi}{2}}}&\gate{X_{-\frac{\pi}{2}}}& \qw\\
\lstick{|0\rangle} &\qw&\qw&\qw&\qw&\qw&\qw&\qw&\ghost{XX_{-\frac{\lambda_2}{2}}}&\gate{X_{\frac{\lambda_2}{2}-\pi}}&\gate{Z_{-\frac{\pi}{2}}}&\gate{X_{-\pi}}& \qw
}
}
        \caption{$Z_4$}
    \end{subfigure}
    \caption{Optimized reduced deuteron VQE ansatz written over a native gate set for trapped-ion quantum computers.}
    \label{fig:nra-deut-opt}
\end{figure}
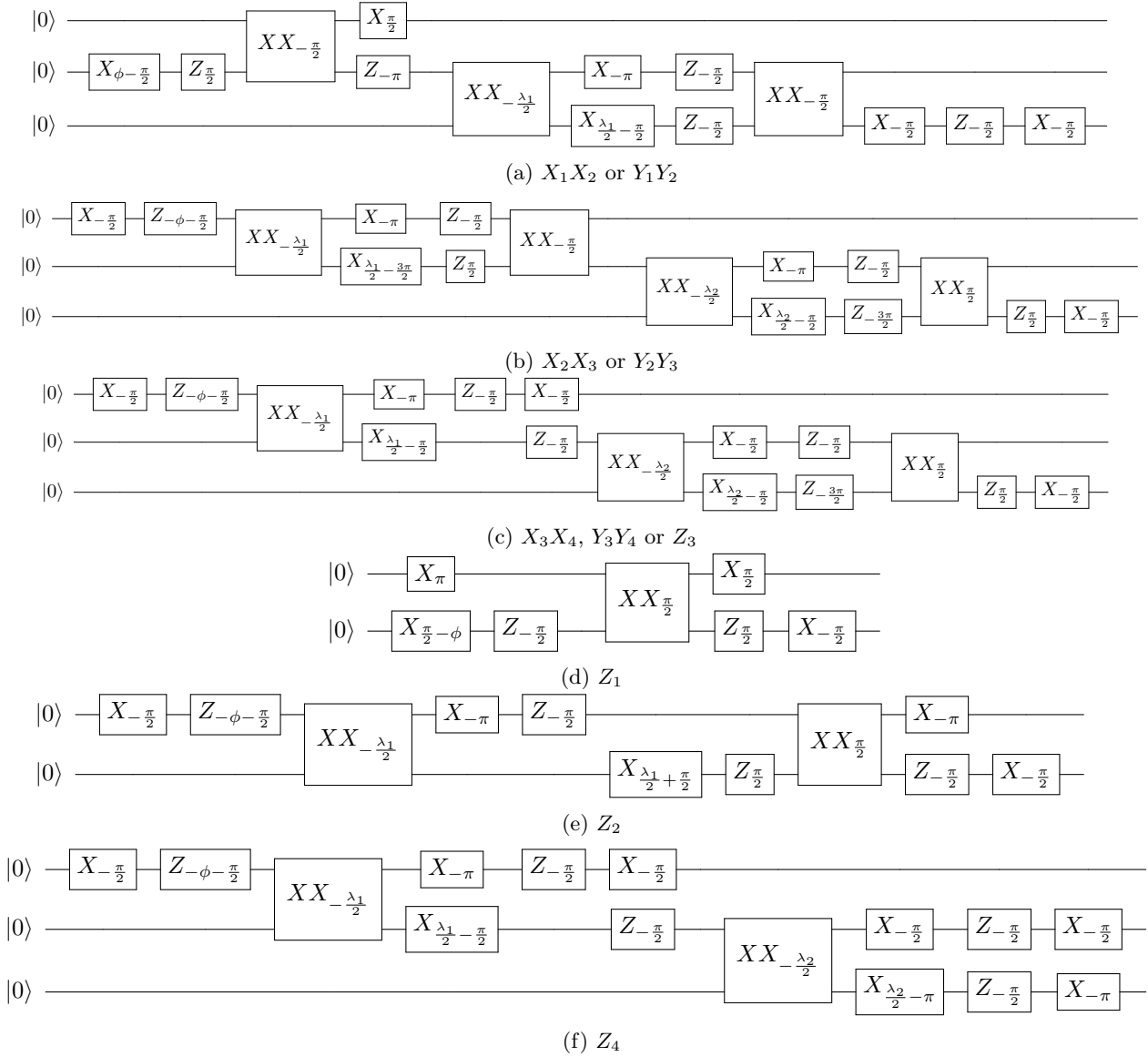

The optimized physical version of the reduced VQE ansatz is given in Figure ~\ref{fig:nra-deut-opt}.

\begin{figure}[H]
\centering
\hspace{0.1cm}\scalebox{0.9}{\Qcircuit @C=1em @R=.7em {
\lstick{|0\rangle}     &  \qw  &   \multigate{1}{XX_{\frac{\gamma_1}{2}}}  &  \qw  &  \qw  &  \qw  &  \qw  &  \qw  &  \qw  &  \qw  &  \qw  &  \qw  &  \qw  &  \qw  &\qw&  \gate{ X_{-\frac{\pi}{2}}}   &\gate{ Z_{-\frac{\pi}{2}}}  &  \gate{X_{\beta -\frac{\pi}{2}}}  & \qw \\ 
\lstick{|0\rangle}     &  \qw  &  \ghost{XX_{\frac{\gamma_1}{2}}}  &  \qw  &  \qw  &   \multigate{1}{XX_{\frac{\gamma_1}{2}}}  &  \qw  &  \qw  &  \qw  &  \qw  &  \qw  &  \qw  &  \qw  &  \qw  &\qw&   \gate{ X_{-\frac{\pi}{2}}}   &\gate{ Z_{-\frac{\pi}{2}}}  &  \gate{X_{\beta -\frac{\pi}{2}}}  & \qw \\ 
\lstick{|0\rangle}    &  \qw  &  \qw  &  \qw  &  \qw  &  \ghost{XX_{\frac{\gamma_1}{2}}}  &  \qw  &  \qw  &   \multigate{1}{XX_{\frac{\gamma_1}{2}}}  &  \qw  &  \qw  &  \qw  &  \qw  &  \qw  &   \multigate{2}{XX_{\frac{\gamma_1}{2}}} & \gate{ X_{-\frac{\pi}{2}}}   &\gate{ Z_{-\frac{\pi}{2}}}  &  \gate{X_{\beta -\frac{\pi}{2}}}  & \qw \\ 
\lstick{|0\rangle}    &  \qw  &  \qw  &  \qw  &  \qw  &  \qw  &  \qw  &  \qw  &  \ghost{XX_{\frac{\gamma_1}{2}}}  &  \qw  &  \qw  &  \multigate{1}{XX_{\frac{\gamma_1}{2}}}  &  \qw  &  \qw &  \ghost{XX_{\frac{\gamma_1}{2}}}  & \gate{ X_{-\frac{\pi}{2}}}   &\gate{ Z_{-\frac{\pi}{2}}}  &  \gate{X_{\beta -\frac{\pi}{2}}}   & \qw \\ 
\lstick{|0\rangle}   &  \qw  &  \qw  &  \qw  &  \qw  &  \qw  &  \qw  &  \qw  &  \qw  &  \qw  &  \qw  &  \ghost{XX_{\frac{\gamma_1}{2}}} &  \qw  &  \qw  &  \ghost{XX_{\frac{\gamma_1}{2}}}  & \gate{ X_{-\frac{\pi}{2}}}   &\gate{ Z_{-\frac{\pi}{2}}}  &  \gate{X_{\beta -\frac{\pi}{2}}}  & \qw
}}
\caption{Optimized QAOA ansatz to compute MAXCUT of the dragon graph when $p = 1$.} \label{fig:dragon-opt}
\end{figure}
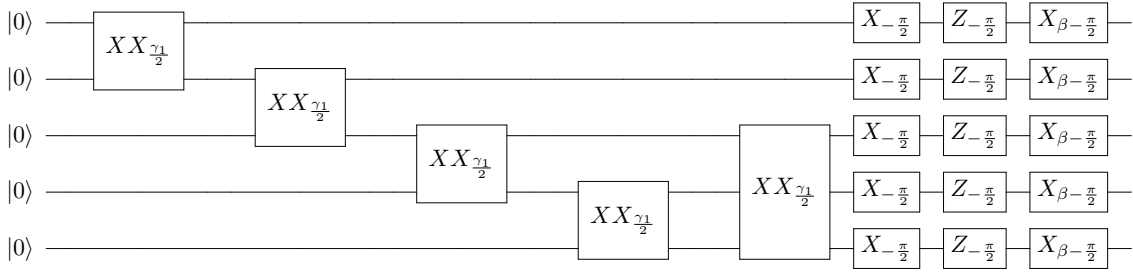

The optimized physical version of the original QAOA ansatz to compute MAXCUT of the dragon graph is given in Figure ~\ref{fig:dragon-opt}.

\begin{figure}[H]
    \begin{subfigure}[b]{\textwidth}
\hspace{0.6cm}\scalebox{0.9}{\Qcircuit @C=1em @R=.7em {
\lstick{|0\rangle}     &  \qw  &   \multigate{1}{XX_{\frac{\gamma_1}{2}}}  &  \qw  &  \qw  &  \qw  &  \qw  &  \qw  &  \qw  &  \qw  &  \qw  &  \qw  &  \qw  &  \qw  &\qw&  \gate{ X_{-\frac{\pi}{2}}}   &\gate{ Z_{-\frac{\pi}{2}}}  &  \gate{X_{\beta -\frac{\pi}{2}}}  & \qw \\ 
\lstick{|0\rangle}     &  \qw  &  \ghost{XX_{\frac{\gamma_1}{2}}}  &  \qw  &  \qw  &   \multigate{1}{XX_{\frac{\gamma_1}{2}}}  &  \qw  &  \qw  &  \qw  &  \qw  &  \qw  &  \qw  &  \qw  &  \qw  &\qw&   \gate{ X_{-\frac{\pi}{2}}}   &\gate{ Z_{-\frac{\pi}{2}}}  &  \gate{X_{\beta -\frac{\pi}{2}}}  & \qw \\ 
\lstick{|0\rangle}    &  \qw  &  \qw  &  \qw  &  \qw  &  \ghost{XX_{\frac{\gamma_1}{2}}}  &  \qw  &  \qw  &   \qw  &  \qw  &  \qw  &  \qw  &  \qw  &  \qw  &  \qw & \gate{ X_{-\frac{\pi}{2}}}   &\gate{ Z_{-\frac{\pi}{2}}}  &  \gate{X_{ -\frac{\pi}{2}}}  & \qw 
}}
\label{figsub:z1z2}
        \caption{$Z_1 Z_2$}
    \end{subfigure}
    
    \begin{subfigure}[b]{\textwidth}
        \centering
\hspace{0.6cm}\scalebox{0.8}{
\Qcircuit @C=1em @R=.7em {
\lstick{|0\rangle}     &  \qw  &   \multigate{1}{XX_{\frac{\gamma_1}{2}}}  &  \qw  &  \qw  &  \qw  &  \qw  &  \qw  &  \qw  &  \qw  &  \qw  &  \qw  &  \qw  &  \qw  &\qw&  \gate{ X_{-\frac{\pi}{2}}}   &\gate{ Z_{-\frac{\pi}{2}}}  &  \gate{X_{ -\frac{\pi}{2}}}  & \qw \\ 
\lstick{|0\rangle}     &  \qw  &  \ghost{XX_{\frac{\gamma_1}{2}}}  &  \qw  &  \qw  &   \multigate{1}{XX_{\frac{\gamma_1}{2}}}  &  \qw  &  \qw  &  \qw  &  \qw  &  \qw  &  \qw  &  \qw  &  \qw  &\qw&   \gate{ X_{-\frac{\pi}{2}}}   &\gate{ Z_{-\frac{\pi}{2}}}  &  \gate{X_{\beta -\frac{\pi}{2}}}  & \qw \\ 
\lstick{|0\rangle}    &  \qw  &  \qw  &  \qw  &  \qw  &  \ghost{XX_{\frac{\gamma_1}{2}}}  &  \qw  &  \qw  &   \multigate{1}{XX_{\frac{\gamma_1}{2}}}  &  \qw  &  \qw  &  \qw  &  \qw  &  \qw  &   \multigate{2}{XX_{\frac{\gamma_1}{2}}} & \gate{ X_{-\frac{\pi}{2}}}   &\gate{ Z_{-\frac{\pi}{2}}}  &  \gate{X_{\beta -\frac{\pi}{2}}}  & \qw \\ 
\lstick{|0\rangle}    &  \qw  &  \qw  &  \qw  &  \qw  &  \qw  &  \qw  &  \qw  &  \ghost{XX_{\frac{\gamma_1}{2}}}  &  \qw  &  \qw  &  \qw  &  \qw  &  \qw &  \ghost{XX_{\frac{\gamma_1}{2}}}  & \gate{ X_{-\frac{\pi}{2}}}   &\gate{ Z_{-\frac{\pi}{2}}}  &  \gate{X_{ -\frac{\pi}{2}}}   & \qw \\ 
\lstick{|0\rangle}   &  \qw  &  \qw  &  \qw  &  \qw  &  \qw  &  \qw  &  \qw  &  \qw  &  \qw  &  \qw  &  \qw &  \qw  &  \qw  &  \ghost{XX_{\frac{\gamma_1}{2}}}  & \gate{ X_{-\frac{\pi}{2}}}   &\gate{ Z_{-\frac{\pi}{2}}}  &  \gate{X_{ -\frac{\pi}{2}}}  & \qw
}
}
        \caption{$Z_2 Z_3$}
    \end{subfigure}
    
        \begin{subfigure}[b]{\textwidth}
\hspace{0.6cm}\scalebox{0.8}{
\Qcircuit @C=1em @R=.7em {
\lstick{|0\rangle}     &  \qw  &  \qw  &  \qw  &  \qw  &   \multigate{1}{XX_{\frac{\gamma_1}{2}}}  &  \qw  &  \qw  &  \qw  &  \qw  &  \qw  &  \qw  &  \qw  &  \qw  &\qw&   \gate{ X_{-\frac{\pi}{2}}}   &\gate{ Z_{-\frac{\pi}{2}}}  &  \gate{X_{ -\frac{\pi}{2}}}  & \qw \\ 
\lstick{|0\rangle}    &  \qw  &  \qw  &  \qw  &  \qw  &  \ghost{XX_{\frac{\gamma_1}{2}}}  &  \qw  &  \qw  &   \multigate{1}{XX_{\frac{\gamma_1}{2}}}  &  \qw  &  \qw  &  \qw  &  \qw  &  \qw  &   \multigate{2}{XX_{\frac{\gamma_1}{2}}} & \gate{ X_{-\frac{\pi}{2}}}   &\gate{ Z_{-\frac{\pi}{2}}}  &  \gate{X_{\beta -\frac{\pi}{2}}}  & \qw \\ 
\lstick{|0\rangle}    &  \qw  &  \qw  &  \qw  &  \qw  &  \qw  &  \qw  &  \qw  &  \ghost{XX_{\frac{\gamma_1}{2}}}  &  \qw  &  \qw  &  \multigate{1}{XX_{\frac{\gamma_1}{2}}}  &  \qw  &  \qw &  \ghost{XX_{\frac{\gamma_1}{2}}}  & \gate{ X_{-\frac{\pi}{2}}}   &\gate{ Z_{-\frac{\pi}{2}}}  &  \gate{X_{\beta -\frac{\pi}{2}}}   & \qw \\ 
\lstick{|0\rangle}   &  \qw  &  \qw  &  \qw  &  \qw  &  \qw  &  \qw  &  \qw  &  \qw  &  \qw  &  \qw  &  \ghost{XX_{\frac{\gamma_1}{2}}} &  \qw  &  \qw  &  \ghost{XX_{\frac{\gamma_1}{2}}}  & \gate{ X_{-\frac{\pi}{2}}}   &\gate{ Z_{-\frac{\pi}{2}}}  &  \gate{X_{ -\frac{\pi}{2}}}  & \qw
}
}
        \caption{$Z_3 Z_4$}
    \end{subfigure}
    
    \begin{subfigure}[b]{\textwidth}
        \centering
\hspace{0.7cm} \scalebox{1}{
\Qcircuit @C=1em @R=.7em {
\lstick{|0\rangle}    &  \qw  &  \qw  &  \qw  &  \qw  &  \qw  &  \qw  &  \qw  &   \multigate{1}{XX_{\frac{\gamma_1}{2}}}  &  \qw  &  \qw  &  \qw  &  \qw  &  \qw  &   \multigate{2}{XX_{\frac{\gamma_1}{2}}} & \gate{ X_{-\frac{\pi}{2}}}   &\gate{ Z_{-\frac{\pi}{2}}}  &  \gate{X_{-\frac{\pi}{2}}}  & \qw \\ 
\lstick{|0\rangle}    &  \qw  &  \qw  &  \qw  &  \qw  &  \qw  &  \qw  &  \qw  &  \ghost{XX_{\frac{\gamma_1}{2}}}  &  \qw  &  \qw  &  \multigate{1}{XX_{\frac{\gamma_1}{2}}}  &  \qw  &  \qw &  \ghost{XX_{\frac{\gamma_1}{2}}}  & \gate{ X_{-\frac{\pi}{2}}}   &\gate{ Z_{-\frac{\pi}{2}}}  &  \gate{X_{\beta -\frac{\pi}{2}}}   & \qw \\ 
\lstick{|0\rangle}   &  \qw  &  \qw  &  \qw  &  \qw  &  \qw  &  \qw  &  \qw  &  \qw  &  \qw  &  \qw  &  \ghost{XX_{\frac{\gamma_1}{2}}} &  \qw  &  \qw  &  \ghost{XX_{\frac{\gamma_1}{2}}}  & \gate{ X_{-\frac{\pi}{2}}}   &\gate{ Z_{-\frac{\pi}{2}}}  &  \gate{X_{\beta -\frac{\pi}{2}}}  & \qw
}
}
        \caption{$Z_4 Z_5$}
    \end{subfigure}
    
        \begin{subfigure}[b]{\textwidth}
\hspace{0.6cm}\scalebox{1}{
\Qcircuit @C=1em @R=.7em {
\lstick{|0\rangle}     &  \qw  &  \qw  &  \qw  &  \qw  &   \multigate{1}{XX_{\frac{\gamma_1}{2}}}  &  \qw  &  \qw  &  \qw  &  \qw  &  \qw  &  \qw  &  \qw  &  \qw  &\qw&   \gate{ X_{-\frac{\pi}{2}}}   &\gate{ Z_{-\frac{\pi}{2}}}  &  \gate{X_{ -\frac{\pi}{2}}}  & \qw \\ 
\lstick{|0\rangle}    &  \qw  &  \qw  &  \qw  &  \qw  &  \ghost{XX_{\frac{\gamma_1}{2}}}  &  \qw  &  \qw  &   \multigate{1}{XX_{\frac{\gamma_1}{2}}}  &  \qw  &  \qw  &  \qw  &  \qw  &  \qw  &   \multigate{2}{XX_{\frac{\gamma_1}{2}}} & \gate{ X_{-\frac{\pi}{2}}}   &\gate{ Z_{-\frac{\pi}{2}}}  &  \gate{X_{\beta -\frac{\pi}{2}}}  & \qw \\ 
\lstick{|0\rangle}    &  \qw  &  \qw  &  \qw  &  \qw  &  \qw  &  \qw  &  \qw  &  \ghost{XX_{\frac{\gamma_1}{2}}}  &  \qw  &  \qw  &  \multigate{1}{XX_{\frac{\gamma_1}{2}}}  &  \qw  &  \qw &  \ghost{XX_{\frac{\gamma_1}{2}}}  & \gate{ X_{-\frac{\pi}{2}}}   &\gate{ Z_{-\frac{\pi}{2}}}  &  \gate{X_{ -\frac{\pi}{2}}}   & \qw \\ 
\lstick{|0\rangle}   &  \qw  &  \qw  &  \qw  &  \qw  &  \qw  &  \qw  &  \qw  &  \qw  &  \qw  &  \qw  &  \ghost{XX_{\frac{\gamma_1}{2}}} &  \qw  &  \qw  &  \ghost{XX_{\frac{\gamma_1}{2}}}  & \gate{ X_{-\frac{\pi}{2}}}   &\gate{ Z_{-\frac{\pi}{2}}}  &  \gate{X_{\beta -\frac{\pi}{2}}}  & \qw
}
}
        \caption{$Z_3 Z_5$}
    \end{subfigure}
    \caption{Optimized  reduced QAOA ansatz written over a native gate set for trapped-ion quantum computers.}
    \label{fig:nra-dg-opt}
\end{figure}
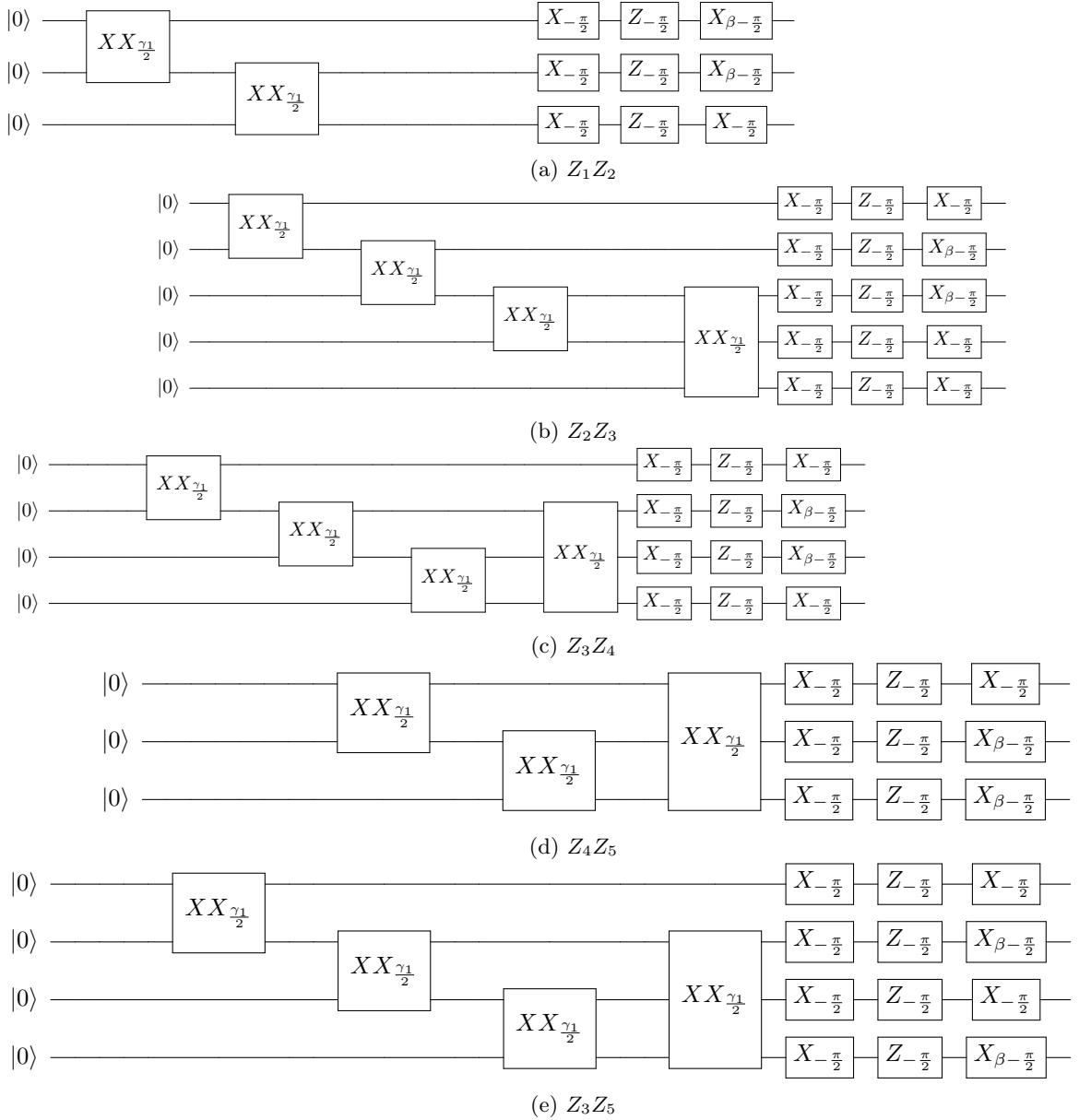

The optimized physical version of the reduced QAOA ansatz is given in Figure ~\ref{fig:nra-dg-opt}.

\section*{Acknowledgements}
This project was conceived during the Aspen Winter Conference, Advances in Quantum Algorithms and Computation 2018 at the Aspen Center for Physics which is supported by National Science Foundation grant PHY-1607611. OS and IK thank the organizers for the hospitality. NML gratefully acknowledges funding by  the PFC@JQI, NSF grant number PHY-1430094. C.H.A. acknowledges financial support from CONACYT doctoral grant No. 
455378.  The authors thank Matthew Keesan, Shantanu Debnath, Vandiver Chaplin, Yunseong Nam, Jungsang Kim, Kenneth Brown, Eugene Dumitrescu, Greg Quiroz, Bill Huggins, and Raphael Pooser for their insightful comments.

\section*{Author Contributions}
Theory was developed by O.S. and I.K; experimental data collected and analyzed by H.N.N. with help from K.L., C.H.A., D.Z. and N.M.L.; O.S. and I.K. performed the circuit design and  in-silico simulation;  and O.S. and I.K. prepared the manuscript with input from all authors.

\section*{Competing Interests}
Provisional patent applications for this work were filed by IonQ, Inc.

\section*{Data availability}
All data needed to evaluate the conclusions are available from the corresponding author upon request.

\section*{Correspondence}
Correspondence and requests for materials should be
addressed to Omar Shehab (email: \texttt{shehab@ionq.co}) or Isaac Kim (email: \texttt{isaac.kim.quantum@gmail.com}).

\bibliographystyle{unsrt}
\bibliography{references}

\appendix

\section{Experimental data for individual terms in the Hamiltonian}
\label{sec:termwiseplot}
In this section, we present the experimental data to show how the expectation values and absolute error converge as the number of measurements increase for individual Hamiltonian term for the reduced ansatz. The result is reported for both the direct and sub-Hamiltonian grouping approach for the RA-VQE ansatz and the direct approach for the RA-QAOA ansatz.  For any observable $O$, $\Delta \langle O\rangle$ is the difference between the experimental and in-silico expectation values.

\begin{figure}[H]
\centering
\includegraphics[scale=0.5]{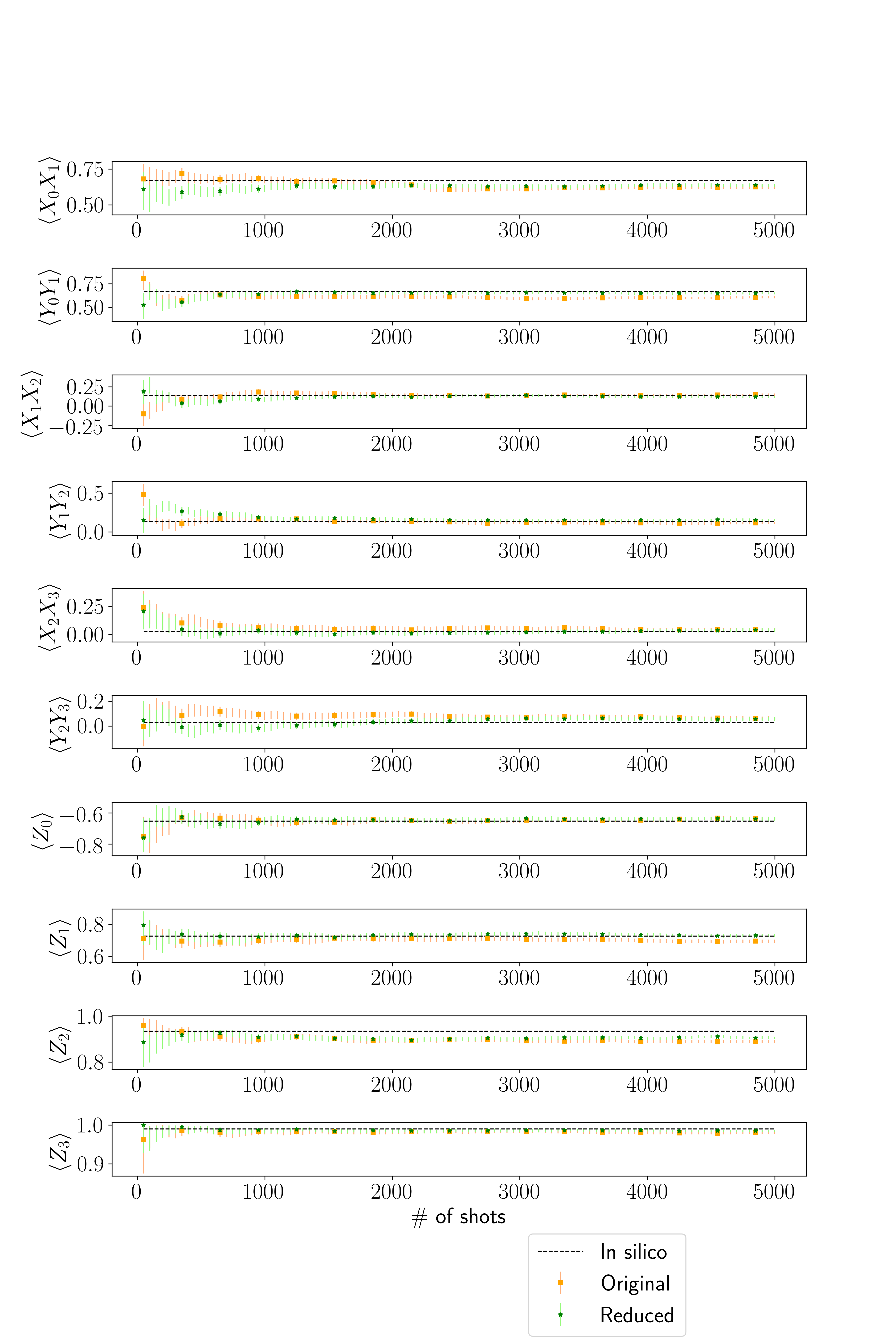}
\caption{Experimental data for the expectation value of individual terms in the deuteron Hamiltonian.}
\label{fig:ind-term-direct}
\end{figure}

\begin{figure}[H]
\centering
\includegraphics[scale=0.45]{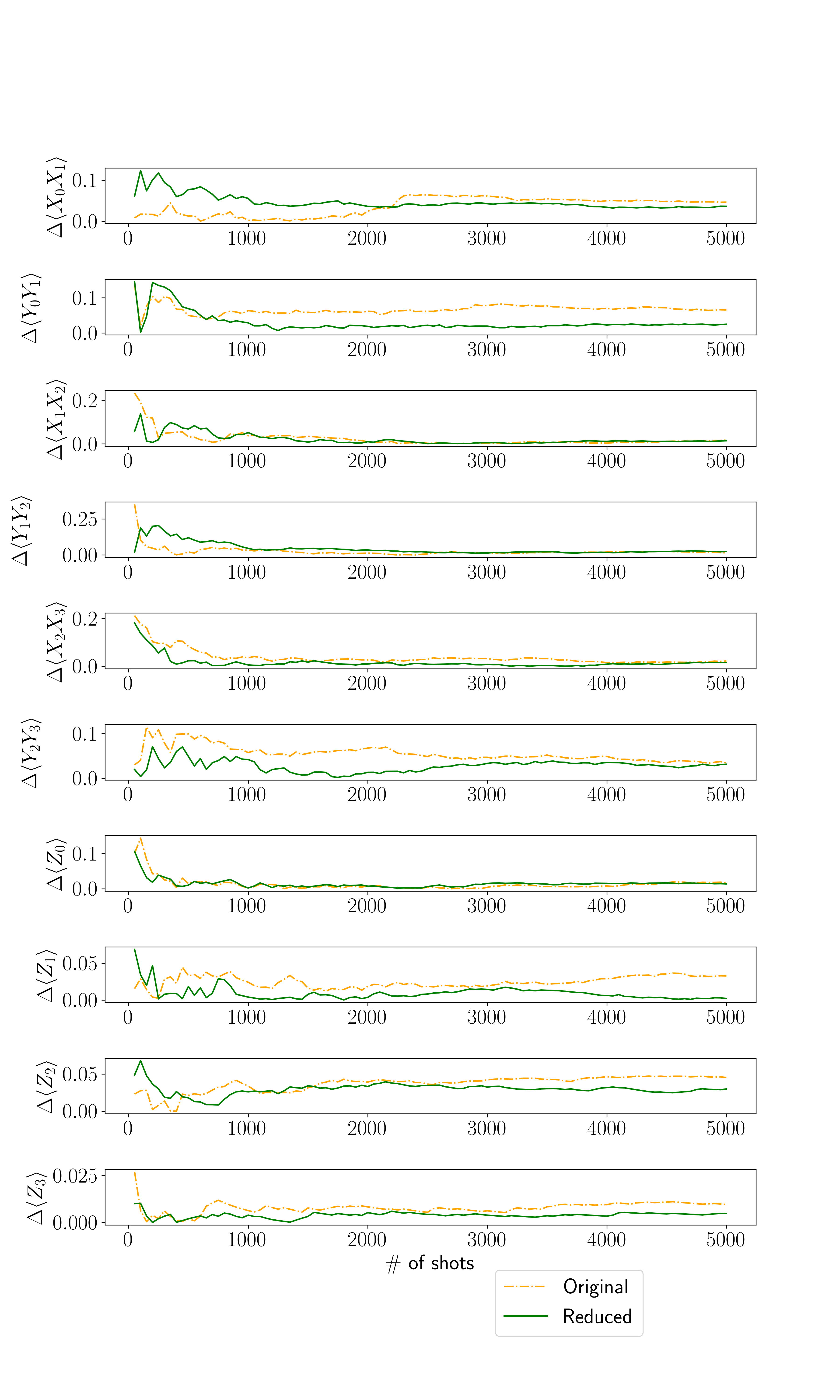}
\caption{Experimental data for the absolute noise in the expectation value of individual terms in the deuteron Hamiltonian.}
\label{fig:ind-term-delta-direct}
\end{figure}

\begin{figure}[H]
\centering
\includegraphics[scale=0.4]{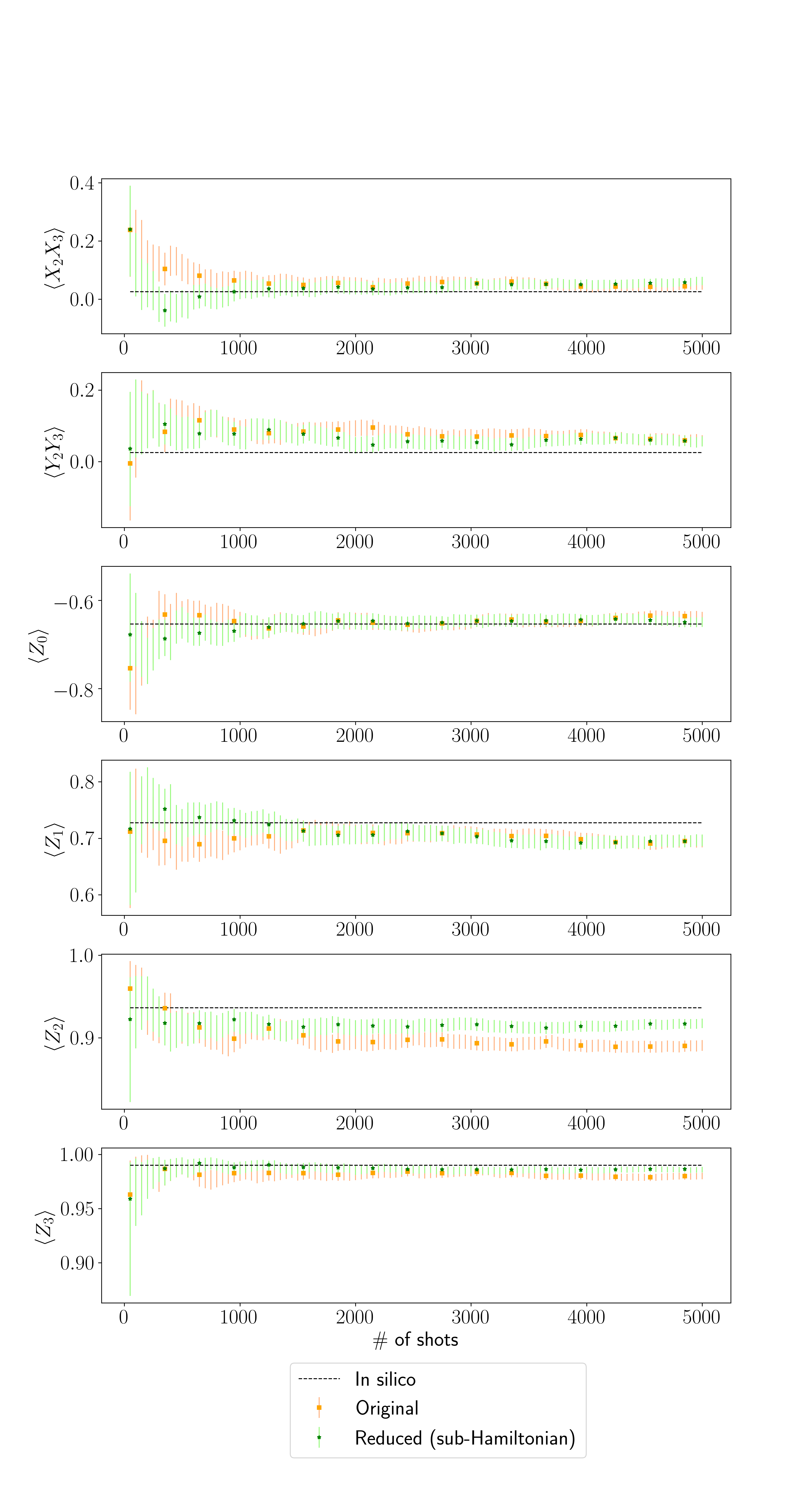}
\caption{Experimental data for the expectation value of individual terms in the deuteron Hamiltonian using sub-Hamiltonian grouping approach.}
\label{fig:ind-term-indirect}
\end{figure}

\begin{figure}[H]
\centering
\includegraphics[scale=0.45]{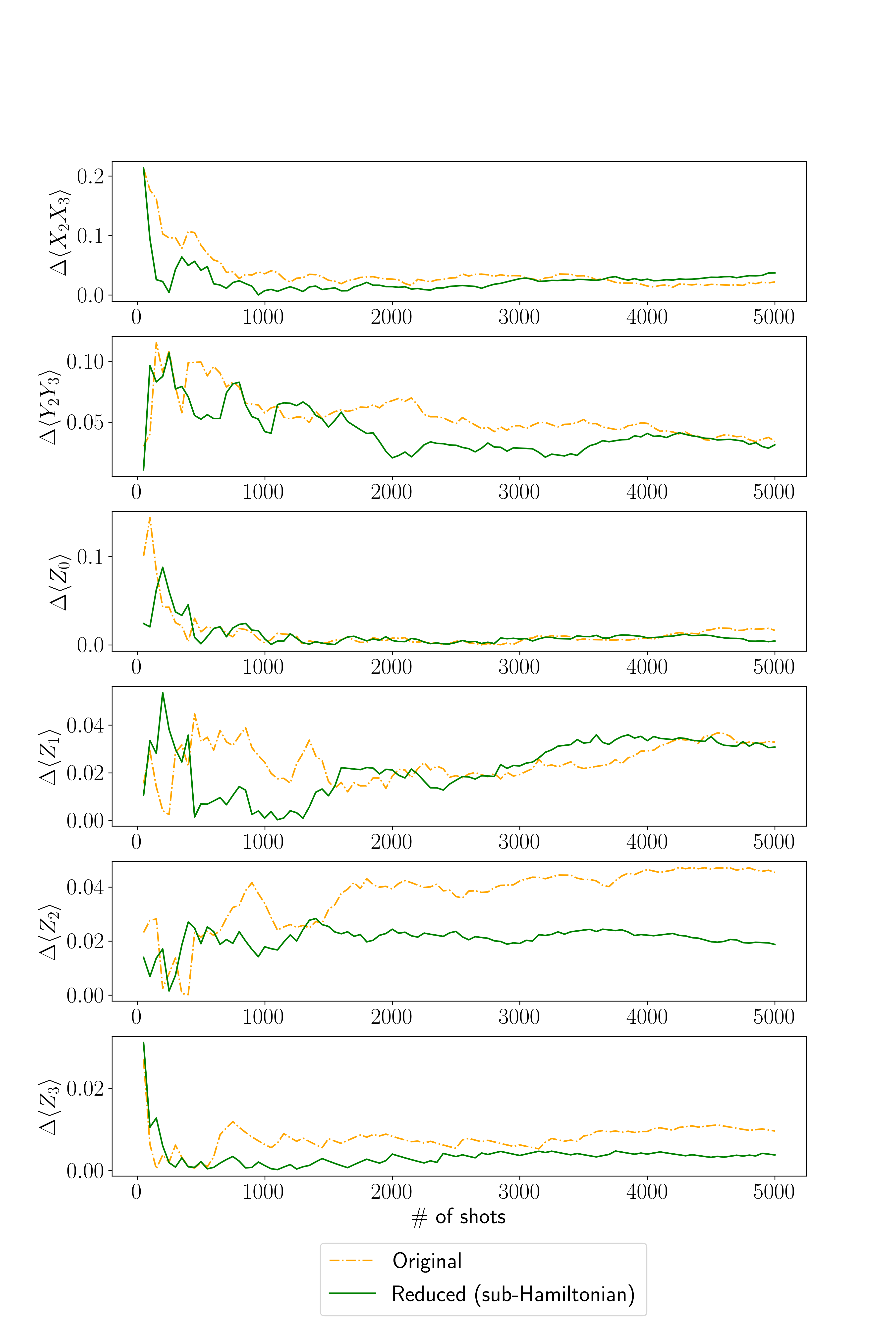}
\caption{Experimental data for the absolute noise in the expectation value of individual terms in the deuteron Hamiltonian using sub-Hamiltonian grouping approach.}
\label{fig:ind-term-delta-direct}
\end{figure}

\begin{figure}[H]
\centering
\includegraphics[scale=0.45]{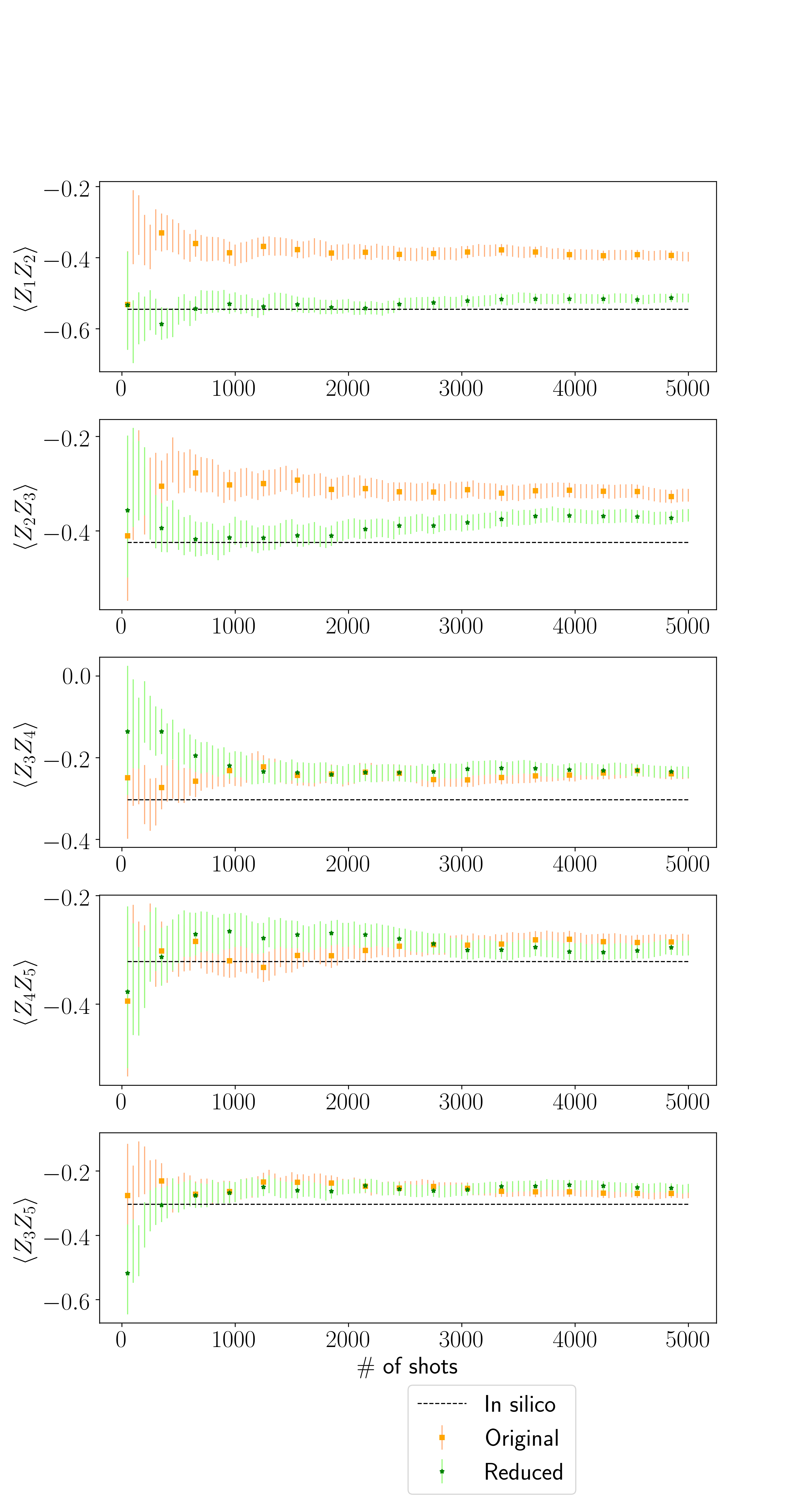}
\caption{Experimental data for the expectation value of individual terms in the dragon graph Hamiltonian.}
\label{fig:ind-term-direct-dragon}
\end{figure}

\begin{figure}[H]
\centering
\includegraphics[scale=0.45]{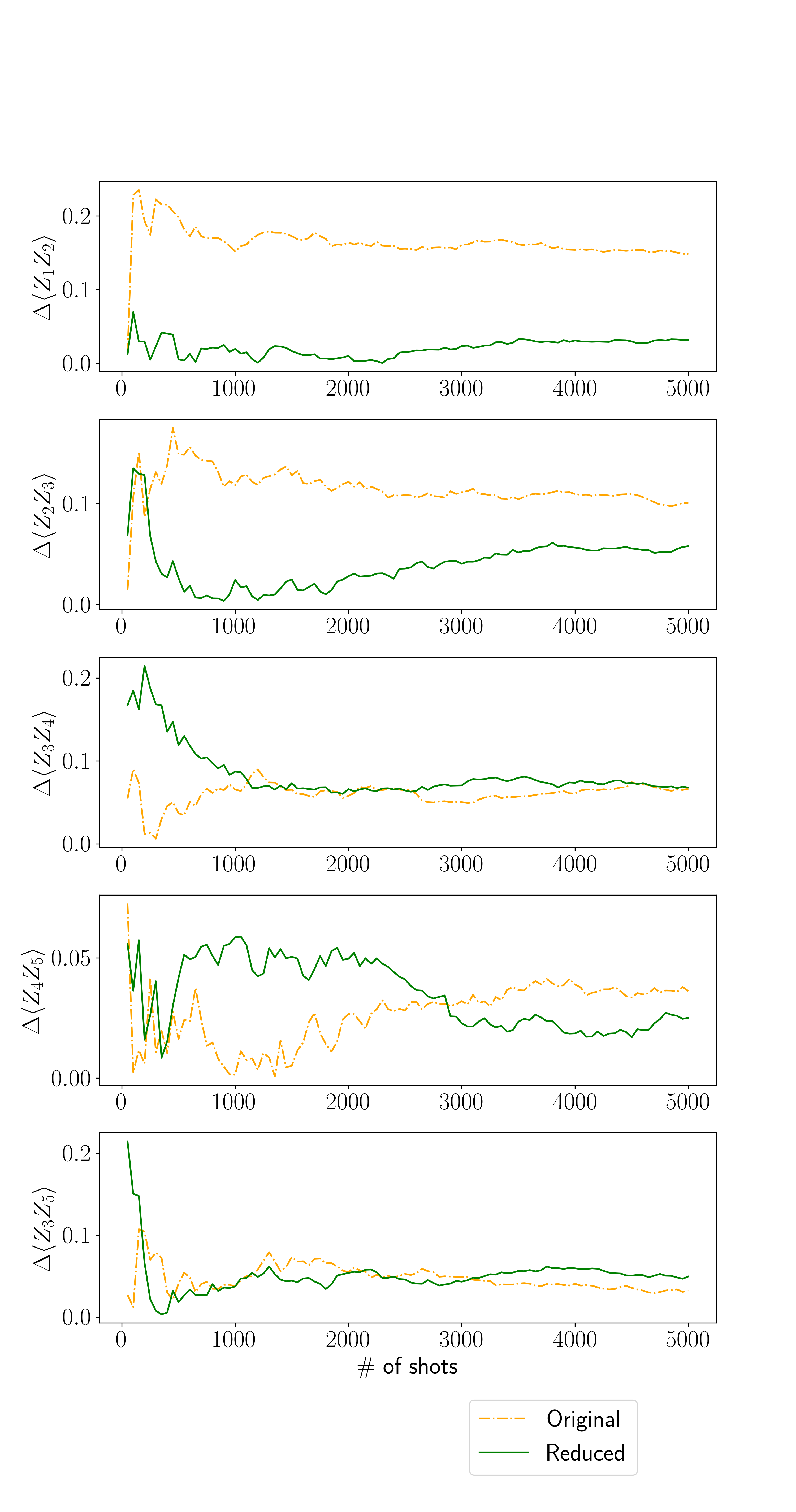}
\caption{Experimental data for the absolute noise in the expectation value of individual terms in the dragon graph Hamiltonian.}
\label{fig:ind-term-delta-direct-dragon}
\end{figure}
\end{document}